\documentclass[12pt,twoside]{article}

\usepackage[
  left=3cm,right=2.5cm,top=3.1cm,bottom=2cm,headheight=2cm,headsep=1.2cm,footskip=1.4cm,twoside
]{geometry}
\usepackage[margin=1cm,font=small]{caption}
\usepackage{amsmath,amssymb}
\usepackage{enumerate}
\usepackage{url}
\usepackage{subfigure}
\usepackage{pstricks}
\newcmykcolor{darkgreen}{1 0 0.6 0.5}  

\usepackage{epsfig}
\usepackage{epsf}
\usepackage{cite}   

\newlength{\nseparation}
\input{babarsymMerge-chka}

\usepackage{cite}   

\newcommand\Amptpbar{\kern 0.18em\overline{\kern -0.18em {\cal A}}_{3\pi}}

\newcommand\Amptpbarkappa{\kern 0.18em\overline{\kern -0.18em A}^{\kappa}{}}
\newcommand\Amptpbarsigma{\kern 0.18em\overline{\kern -0.18em A}^{\sigma}{}}

\newcommand\Nbpm{{\kern 0.18em\overline{\kern -0.18em N}}^{+-}}
\newcommand\Nbmp{{\kern 0.18em\overline{\kern -0.18em N}}^{-+}}

\newcommand\BRpmb{{\cal \kern 0.18em\overline{\kern -0.18em  B}}{}_{\rho\pi}^{+-}}
\newcommand\BRmpb{{\cal \kern 0.18em\overline{\kern -0.18em  B}}{}_{\rho\pi}^{-+}}

\newcommand\BRipmb{{\cal \kern 0.18em\overline{\kern -0.18em  B}}{}_{\rho^+\pi^-}}
\newcommand\BRimpb{{\cal \kern 0.18em\overline{\kern -0.18em  B}}{}_{\rho^-\pi^+}}

\newcommand\Abar{\kern 0.18em\overline{\kern -0.18em A}{}}

\begin{document}
\begin{titlepage}
  {\raggedleft 
  }
  \vskip 2em
  {\centering
    \large 
    M.~Ehlert$^{a}$, A.~Hollnagel$^{b}$, I.~Korol$^{a}$, A.~Korzenev$^{c}$, H.~Lacker$^{a}$, 
    P.~Mermod$^{c}$, J.~Schliwinski$^{a}$, L.~Shihora$^{a}$, P.~Venkova$^{a}$, M.~Wurm$^{b}$\\
  }
  \vskip 2em
  {\centering {\Large Proof-of-principle measurements with a liquid-scintillator detector using
  wavelength-shifting optical modules}}
  \vskip 2.2em
  {
    \noindent
    Based on test-beam measurements, we study the response of a liquid-scintillator detector equipped with 
    wavelength-shifting optical modules, that are proposed e.g. for the IceCube experiment and the SHiP experiment, 
    and adiabatic light guides that are viewed either by a photomultiplier tube or by an array of silicon photomultipliers. 
    We report on the efficiency, the time resolution and the detector response to different particle types and point out 
    potential ways to improve the detector performance.
   }
    \vskip 1.5em \noindent
{\small \em $^{a}$ Humboldt-Universit\"at zu Berlin,
                   Institut f\"ur Physik,
                   Newtonstr. 15,
                   D-12489 Berlin, Germany,
                   {e-mail: lacker@physik.hu-berlin.de}}\\[0.2cm] 
{\small \em $^{b}$  Johannes Gutenberg-Universit\"at,
                   Institut f\"ur Physik,
                   Staudingerweg 7,
                   D-55128 Mainz, Germany}
\\[0.2cm] 
{\small \em $^{c}$  Universit\'e de Gen\`eve,
                   Departement de Physique Nucleaire et Corpusculaire,
                   CH-1211 Gen\`{e}ve, Switzerland}
\\[0.2cm] 
\vskip1em

\end{titlepage}


\noindent

\section{Introduction}
\label{sec:introduction}
Large-volume liquid-scintillator detectors (LSDs) are often used in neutrino-physics experiments, 
either for neutrino detection or as a veto detector. For detection of the scintillation photons, 
large-area spherical photomultiplier tubes (PMTs) are typically employed. 
In this paper, we explore the possibility to read out such large-volume LSDs with Wavelength-shifting 
Optical Modules (WOMs). This technology might be used in future neutrino-physics experiments or for the 
Surrounding Background Tagger of the proposed SHiP experiment~\cite{Anelli:2015pba} at the CERN SPS. 
WOMs as large-area photodetectors of low cost and low noise were proposed first for an upgrade of the 
IceCube experiment~\cite{WOMs}. They are considered as a promising replacement for the large-area 
spherical PMTs.

\begin{figure}
     \centering
     \subfigure[]{
          \label{fig:WOM_Principle}
             \includegraphics[scale=0.35]{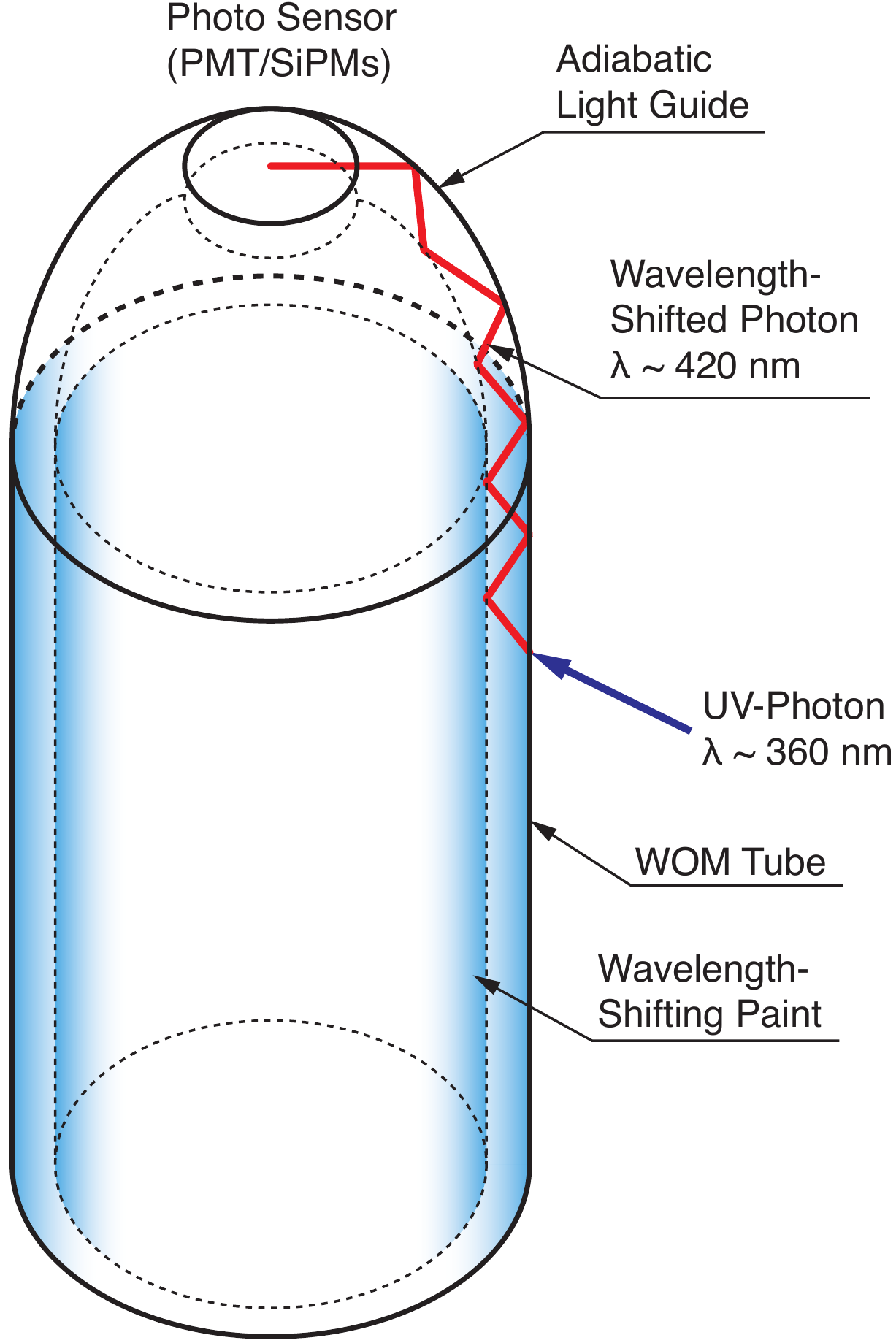}}
     \subfigure[]{
          \label{fig:EmissionAbsorptionSensitivities}
             \includegraphics[scale=0.35]{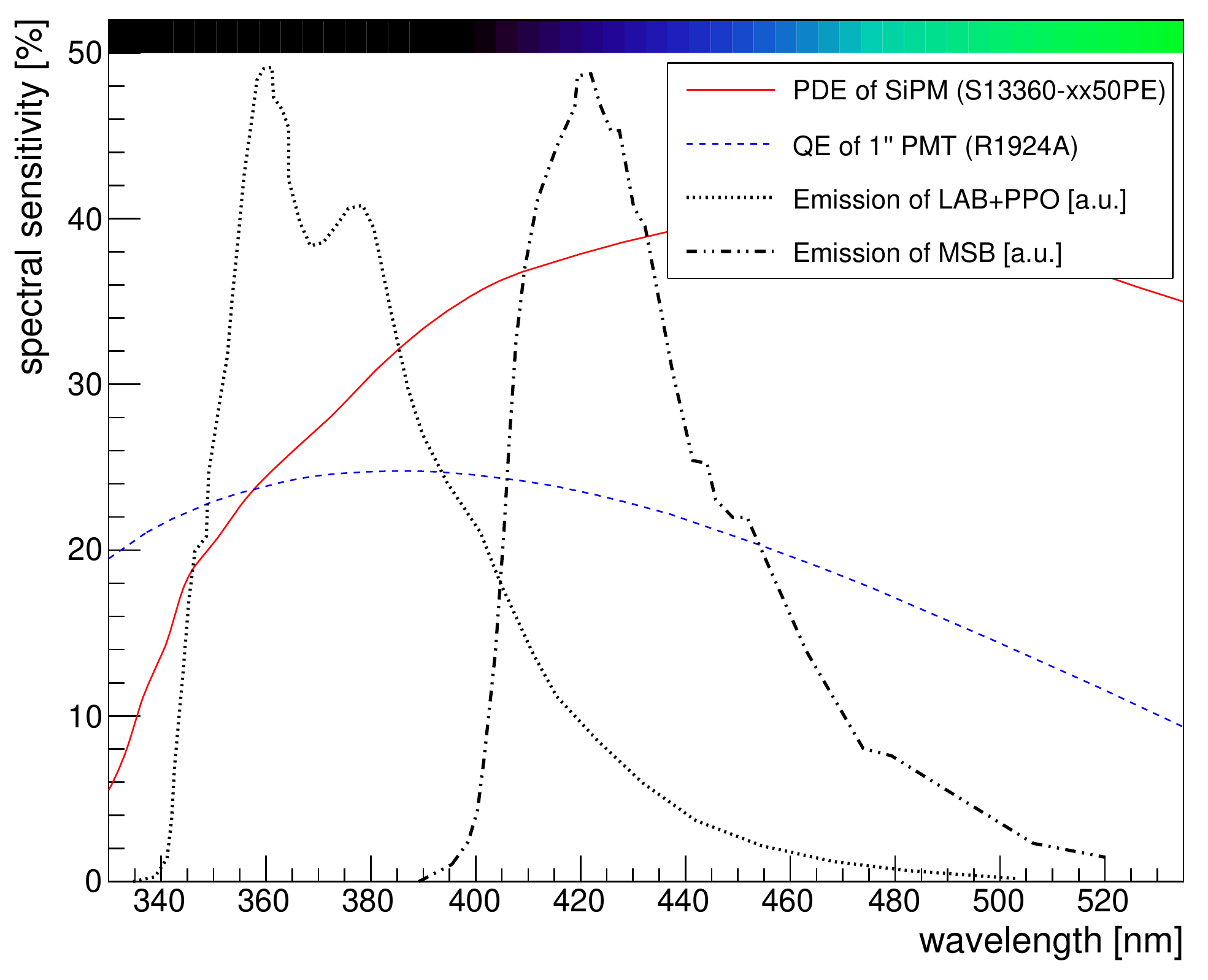}}
  \caption{(a) The working principle of a wavelength-shifting optical module (WOM).
(b) Emission spectra of the liquid scintillator (LAB+PPO) and the wavelength-shifting
paint used on the WOMs (Bis-MSB). Also shown is the quantum effiency (QE) of
the photomultiplier and the photo detection efficiency of the SiPMs, at 3~V as quoted by 
the producer company Hamamatsu photonics, coupled to the WOMs in this setup.}
  \label{fig:WOM_PrincipleSpectra}
\end{figure}

A WOM consists of a cylinder or tube, typically made out of quartz glass or polymethyl methacrylate 
(PMMA), the surface of which is coated with a wavelength-shifting paint. The WOM is placed inside a 
UV-transparent cylindrical vessel, which is of larger diameter than the WOM and separates it from 
the detector material, that can be either a (liquid) scintillator or water/ice to detect Cherenkov photons. 
If a UV photon hits the WOM, it is absorbed inside the wavelength-shifting paint and subsequently
a secondary photon of longer wavelength in the range of visible light is emitted isotropically. 
The WOM is designed to act as a light guide providing total reflection at the glass-air interface.
Neglecting transport losses, the probability that a secondary photon is transported via total reflection 
to one of either ends of the WOM, where it can then be detected by a photosensor, is about 
$75~\%$~\cite{WOMs}. If only one end of the WOM is instrumented with a photosensor and the other end 
is open, a part of the photons being transported to the uninstrumented end will also be totally reflected 
there and eventually detected at the instrumented WOM end. The detection principle is depicted in 
Fig.~\ref{fig:WOM_Principle}.
The photosensor only needs to cover the exit area of the WOM tube or cylinder. 
The gain in effective detection area with respect to the photosensor area is thus mainly
defined by the WOM length. For a WOM tube, the detection area can be further increased 
by enlarging the tube diameter and guiding the secondary light exiting at the end of the 
WOM tube via a light guide to a photosensor. 
Using low-radioactivity quartz glass or PMMA and given the small photosensor
area, the noise of such a large-area WOM is expected to be low.

In this paper, we report on proof-of-principle measurements with a LSD that is viewed by 
WOM tubes transporting the light via a light guide either to a photomultipler tube (PMT) 
or an array of large-area silicon photomultipliers (SiPMs). The measurements were performed 
at the CERN SPS test-beam facility using different particle types and momenta.

\section{Liquid-Scintillator Detector and WOMs}
\label{sec:liquidscintillatordetector}
A box with a volume of $50 \times 50 \times 25~{\rm cm}^3$ made out of 1 cm thick 
black-colored ABS plastic plates glued together is filled with a liquid scintillator 
consisting of 
linear alkylbenzene (LAB)~\footnote{linear alkylbenzene (type Hyblene 113) produced by Sasol Italy S.p.A.} 
as a solvent and 1.5~g/l Diphenyloxazole 
(PPO)~\footnote{2,5-Diphenyloxazole (ca. $99~\%$, for scintillation) produced by Carl Roth GmbH + Co. KG Schoemperlenstr. 1-5, 76185 Karlsruhe, Germany} as a fluorescent. 
These liquid-scintillator constituents are used in state-of-the-art large-volume $\nu$ 
experiments as SNO+~\cite{OKeeffe:2011dex} and JUNO~\cite{Djurcic:2015vqa}.
PPO emits in the wavelength range between about 330~nm and 450~nm with the emission maximum at 
about 360~nm (see Fig.~\ref{fig:EmissionAbsorptionSensitivities}). Laboratory measurements have 
shown that the absorption length of this liquid scintillator is about 1~m at 380~nm and 0.5~m 
at 360~nm~\cite{Hackbarth}. To avoid detector performance deterioration induced by oxygen, 
the LSD is constantly flushed with dry nitrogen gas. \\

\begin{figure}
  \begin{center}
             \includegraphics[scale=0.50]{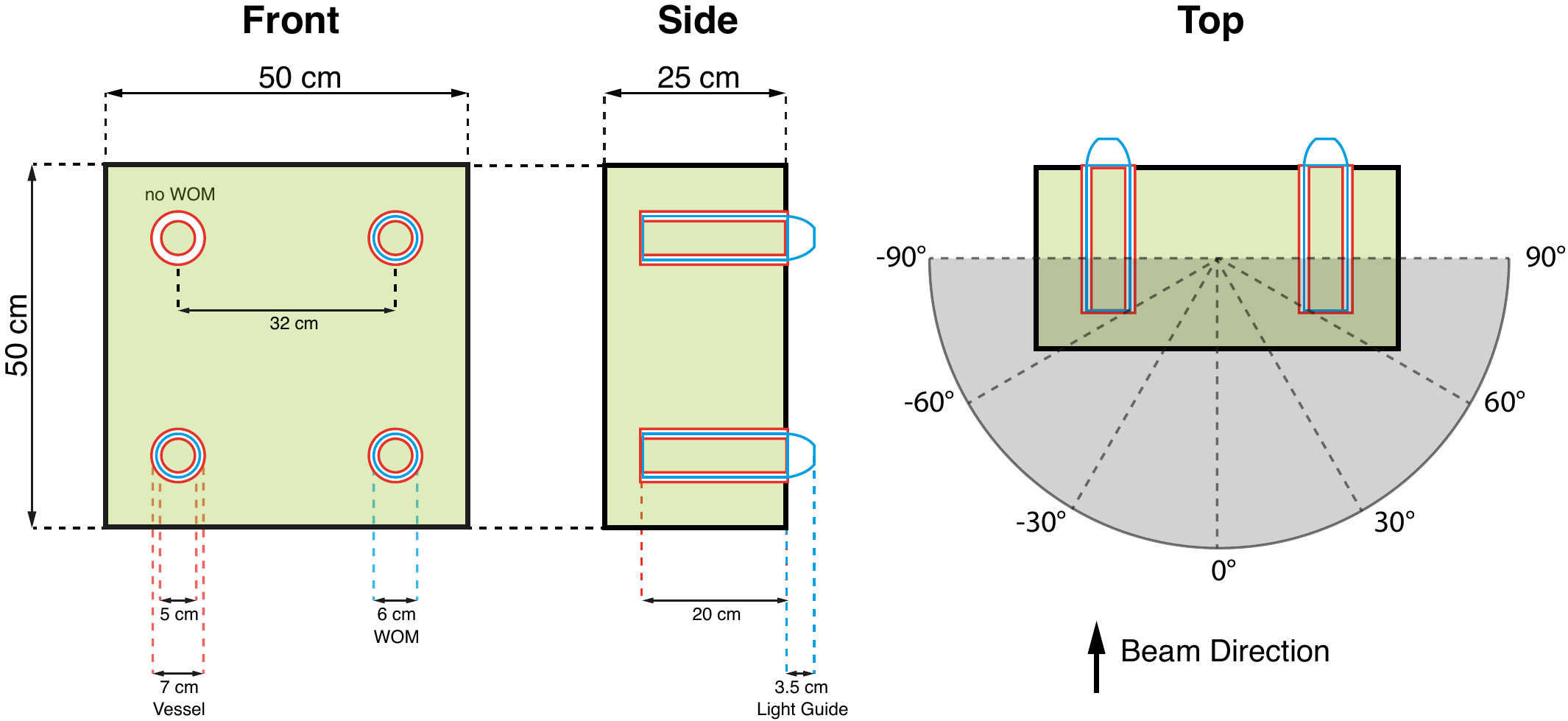}
  \end{center}
  \caption{A sketch of the liquid-scintillator detector module used for the test-beam measurements. 
(Left) front view, (middle) side view, (right) top view showing the definition of the beam incident 
angle in the testbeam measurements.}
  \label{fig:lsbox}
\end{figure}
\begin{figure}
     \centering
     \subfigure[]{
          \label{fig:PMMA_Vessel}
             \includegraphics[scale=0.50]{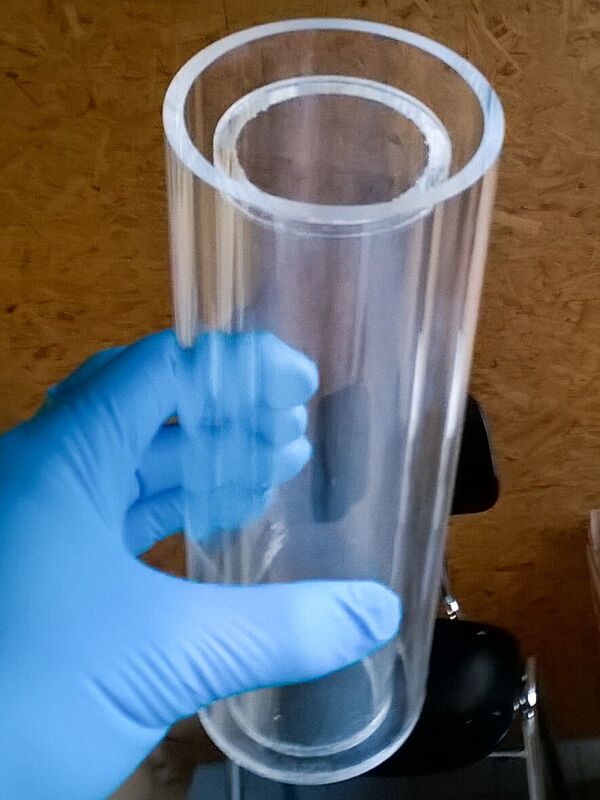}}
     \subfigure[]{
          \label{fig:LightGuide_WOM}
             \includegraphics[scale=0.50]{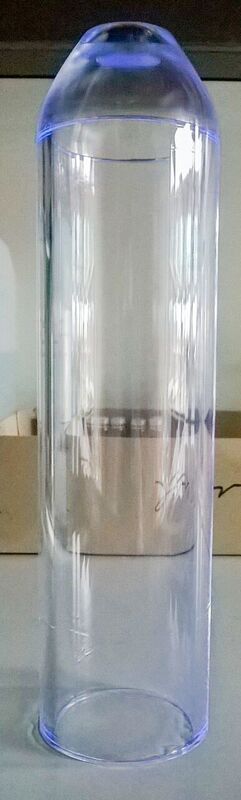}}
  \caption{(a) The double-walled PMMA vessel to separate the WOM from the liquid scintillator.
(b) The light guide glued to the dip-coated WOM tube to be placed in the PMMA vessel.}
  \label{fig:PMMA_Vessel2_LightGuide_WOM}
\end{figure}

A sketch of the liquid-scintillator detector module is shown in Fig.~\ref{fig:lsbox}.
The inner sides of the front and back covers is lined with 
Tyvek~\footnote{Tyvek foil, registered trademark of the DuPont company} foil~\cite{Bellini:2011yd}
to allow diffuse reflection of UV photons. At four positions of the module cover plate, holes 
allow the installation of the WOMs. For the testbeam measurements, the LSD was equipped with three WOMs.
To enable total reflection at its surface, the WOM needs to be surrounded by air. It is thus placed 
inside a 20cm long UV-transparent PMMA vessel which is glued to the module cover and separates the WOM 
from the liquid scintillator. Fig.~\ref{fig:PMMA_Vessel} shows 
a photo of a PMMA vessel~\cite{MaximilianEhlert-Bachelorarbeit}. This PMMA vessel is made 
out of two tubes. The inner tube is sealed against the liquid scintillator via a glued-on disk of 4mm-thick PMMA, 
and both tubes are connected on the opposite side by a ring of the same material.
This geometrical shape of the PMMA vessel with an outer and inner PMMA tube allows one to have also 
liquid scintillator in between the inner walls of the PMMA vessel, increasing the photon detection 
efficiency for particles traversing the module at the location of the WOM. \\

The WOMs themselves are made out of 20~cm long PMMA tubes of 6~cm diameter, with a wall thickness of 4~mm.
They are dip-coated on both sides with a wavelength-shifting paint developed by D.~Hebeker~\cite{WOM-Masterarbeit-DustinHebeker} 
that consists of Toluene (100~ml) as a solvent, Paraloid B72 (25~g) as a binder, 
and the two wavelength shifters Bis-MSB (0.15~g) and p-Terphenyl (0.3~g). 
The solvent is heated to just below the boiling point in order to solve binder and
WLS, which are added in small quantities.
The utilised wavelength shifters absorb photons with high probability in the range 
between 300~nm and 400~nm and emit secondary photons in the range between 400~nm and 520~nm 
with the emission maxiumum at 420~nm (see Fig.~\ref{fig:EmissionAbsorptionSensitivities}). 
Bis-MSB is spectrally particularly well-matched to absorb the 
PPO emission spectrum~\cite{Djurcic:2015vqa}.\\

Using acetone, one end of the WOM tube is fused onto a 3.5~cm-long light guide made out of PMMA
A~\footnote{produced by K\"umpel Kunststoff-Verarbeitungs GmbH}. 
Its shape was optimised for minimal light losses while mapping the WOM ring shape onto a disc of 
2.8~cm diameter~\cite{Lightguide-Bachelorarbeit}. Fig.~\ref{fig:LightGuide_WOM} 
shows the light guide and WOM tube glued together~\cite{MaximilianEhlert-Bachelorarbeit}.\\

\begin{figure}
  \begin{center}
             \includegraphics[scale=0.25]{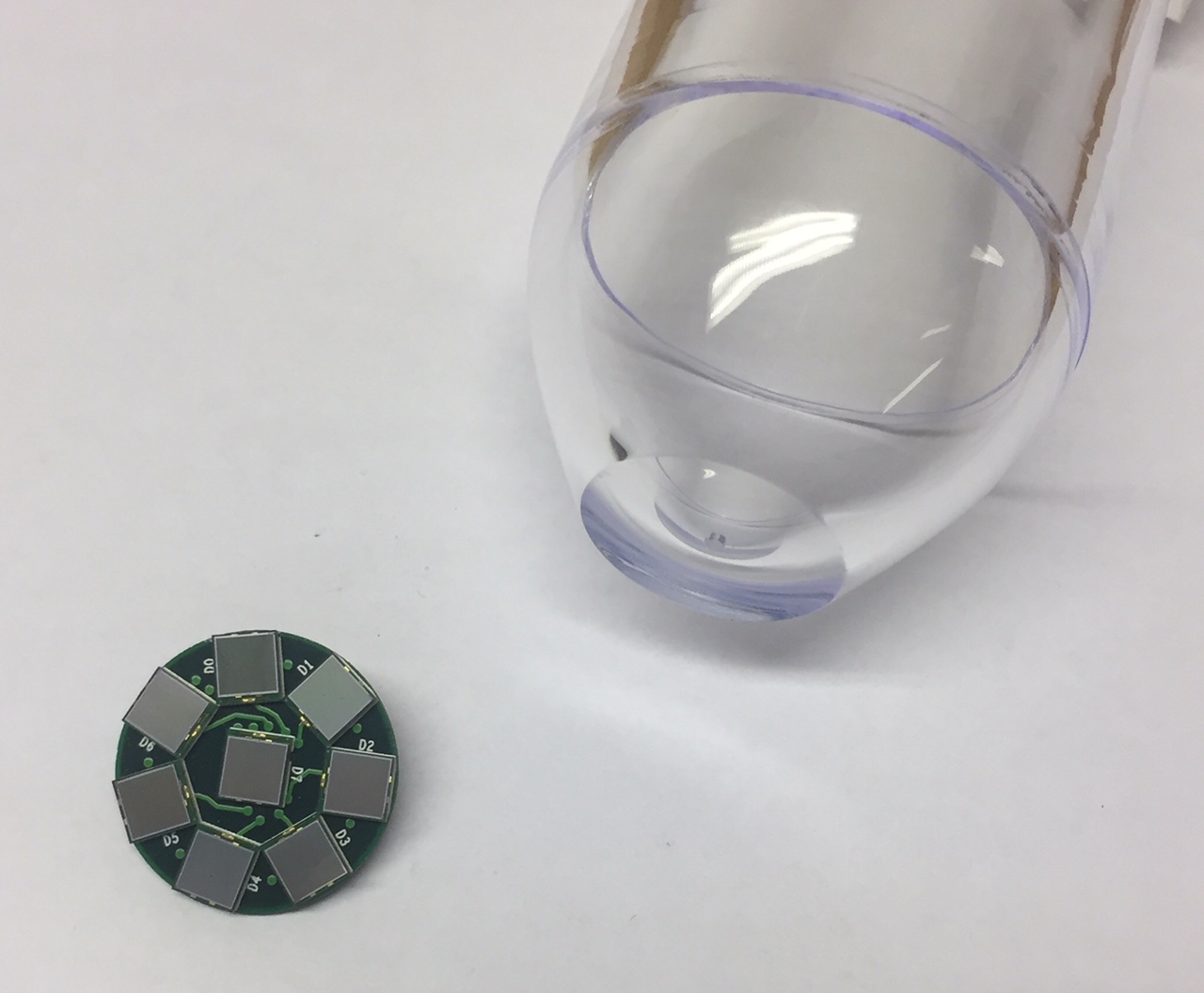}
  \end{center}
  \caption{The eight SiPMs, each of an area of $6 \times 6~{\rm mm}^2$, mounted on the PCB,
which is placed on the exit of the WOM light guide (visible in the right-hand side of the picture).}
  \label{fig:SiPM_PCB}
\end{figure}

During the test-beam measurements, two WOMs were viewed by a Hamamatsu R1924A PMT,
called PMT1 and PMT2, with a photocathode diameter of 2.3~cm, thus only covering 
the inner 67.5~\% of the light guide exit. 
A third WOM was instrumented with an array of eight Hamamatsu S13360-6025PE SiPMs, 
each of which covering an area of $6 \times 6~{\rm mm}^2$. They are placed on a PCB designed 
by the University of Geneva, as shown in Fig.~\ref{fig:SiPM_PCB}, where one SiPM is positioned 
in the centre of the PCB and the other ones around it. The diameter of the PCB is 3~cm and the 
SiPMs cover $40~\%$ of the light guide exit. While the SiPMs thus cover a smaller fraction of 
the light guide exit than the Hamamatsu R1924A PMT, they are partly placed on its outer regions.  
Both, the PMTs as well as the SiPM array were optically coupled to the light guide exit
with a silicone grease~\footnote{Baysilone-paste (medium viscosity) produced by Bayer}.
During the test beam, the SiPMs were operated with a bias voltage of 59.2~V.
The SiPM array was read out by an eight-channel MUSIC\,R1 ASIC~\cite{MUSIC-ASIC} 
providing individual channel read out as well as the analog sum of all eight channel signals.
This approach was already approbated earlier for the readout 
of a plastic scintillator bar~\cite{Betancourt:2017sex}. 

\section{Testbeam setup and Data taking}
\label{sec:testbeamsetup}
Data were taken for three days in September 2017 with beams of hadrons, muons or electrons 
at the CERN SPS north area beamline T2 H2. The SPS provided spills every 13~s or 36~s with 
spill lengths varying between 4~s and 5.5~s.  
For the secondary hadron and muon beams, a momentum of 150~GeV/c was chosen, while
the momentum for the tertiary electron beam was 20~GeV/c. 
Depending on the particle type, typical particle rates during a spill varied between 0.1 and 1~kHz 
and the typical diameter of the beam at the position of the LSD module was on the order of a few cm.\\

\begin{figure}
  \begin{center}
             \includegraphics[scale=0.50]{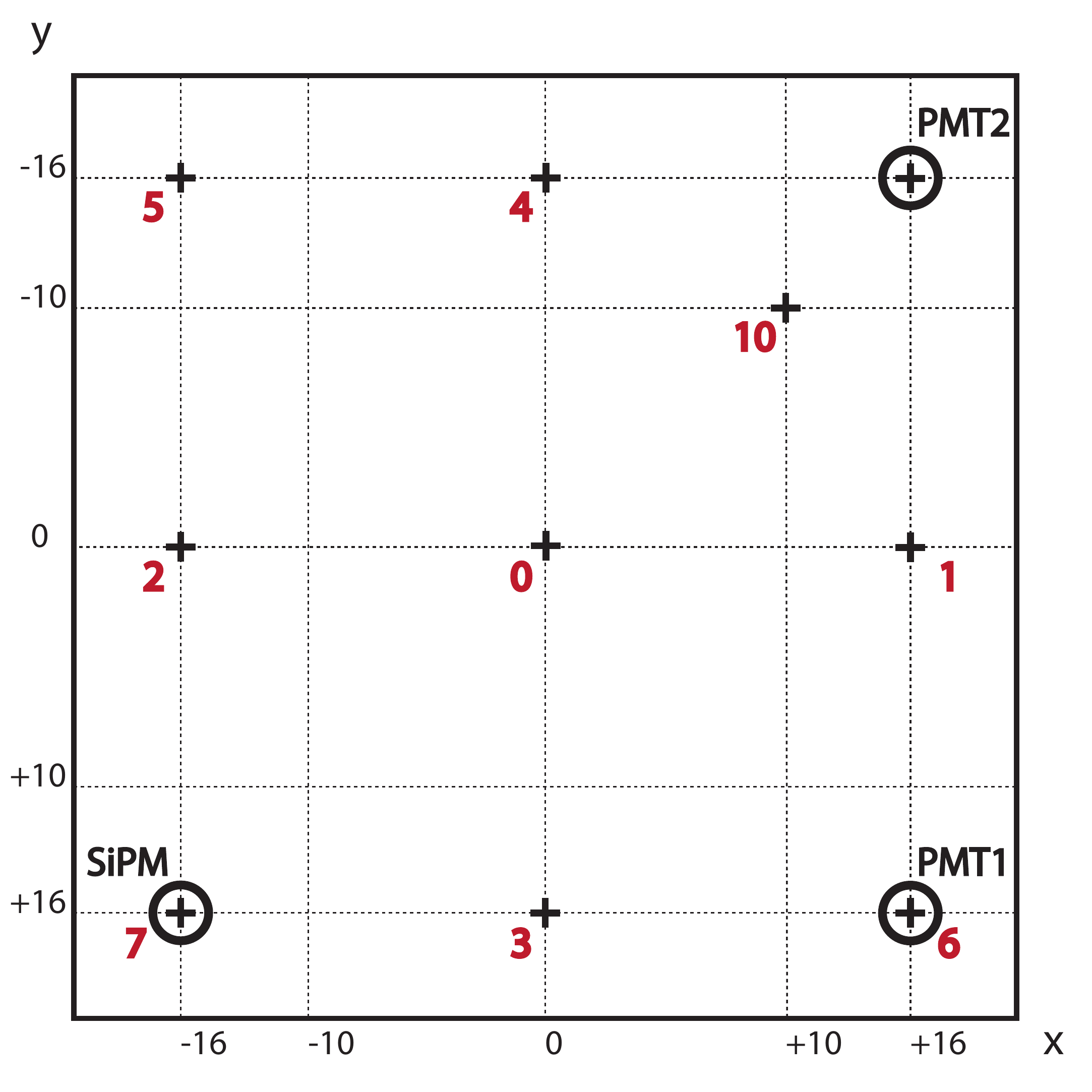}
  \end{center}
  \caption{Sketch of the LSD module indicating the beam positions on the detectors with reference numbers.
The WOMs equipped either by a PMT or by a SiPM array are labeled accordingly.}
  \label{fig:Testbeam_Box_Scetch}
\end{figure}

A 'DESY table' allowed for horizontal and vertical movement of the detector module, 
enabling measurements with the beam spot at several reference points.
Fig.~\ref{fig:Testbeam_Box_Scetch} shows a 
sketch of the module indicating these positions as well as the positions of the WOMs with a label 
indicating whether a PMT or the SiPM array was used. In addition, the module could be rotated around 
its vertical symmetry axis to change the incident angle $\theta_{inc}$ of the particles with 
respect to the front and back covers of the LSD module, with $\theta_{inc}=0^{\circ}$ (see Fig.~{\ref{fig:lsbox}}, right-hand plot) 
denoting the default position where the beam was perpendicular to the detector module cover.\\

\begin{figure}
  \begin{center}
             \includegraphics[scale=0.50]{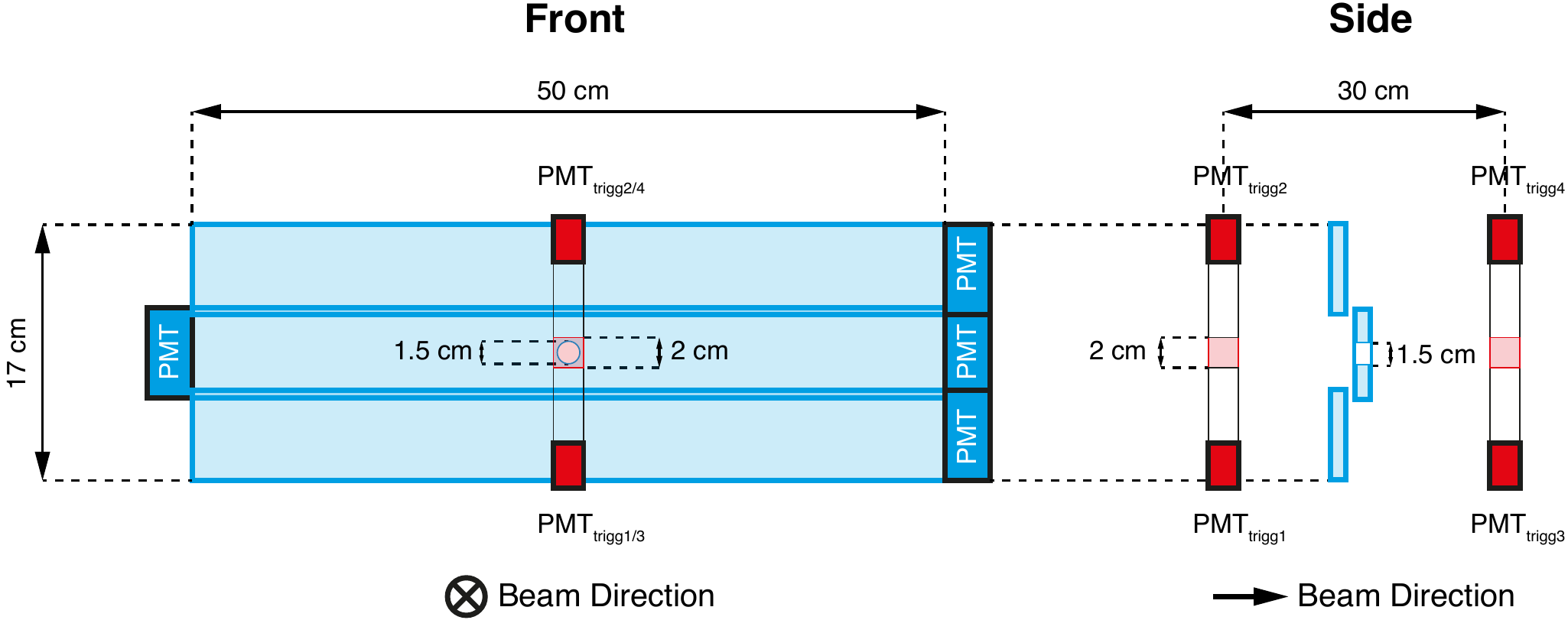}
  \end{center}
  \caption{Sketch of the plastic-scintillator trigger-and-veto system located 75~cm upstream of the LSD module.}
  \label{fig:Trigger_Veto}
\end{figure}

Two plastic scintillators (EJ-228) with dimensions $2 \times 2 \times 2~{\rm cm}^3$ acted as 
trigger and were placed 0.75~m upstream of the LSD module. The scintillators were viewed from both 
sides by Hamamatsu R4998 PMTs via 5~cm long light guides, hereafter referred to as $PMT_{trigg1}$ 
and $PMT_{trigg2}$, respectively, $PMT_{trigg3}$ and $PMT_{trigg4}$. 
Three veto scintillators were installed between the two trigger scintillators to suppress 
events with additional particles not passing the trigger scintillators. 
Each veto scintillator was 50~cm long in horizontal direction, 6~cm in vertical direction, 
and had 1 cm thickness. 
The veto scintillators were sligthly overlapping in vertical direction so that they covered 
an area of 50 cm $\times$ 17~cm. The central veto scintillator had a hole of 1.5~cm diameter, 
thus defining the accepted position of the beam spot on the LSD module. 
Fig.~ \ref{fig:Trigger_Veto} shows a sketch of the plastic scintillator trigger-and-veto system.
All signals from the trigger and veto scintillators, the PMTs from two WOMs and 
and SiPM signals of seven out of eight individual channels and the sum-channel from 
the other WOM were sent to a 16-channel WaveCatcher module~\cite{WaveCatcher} for data acquisition.
The module was operated at the 3.2~GS/s sampling rate, thus covering a 320~ns time window.
Event triggers were only accepted if all four trigger PMTs fired and no signal 
from the veto counters was present within a time window of~30 ns. 
Data taken with constant conditions were grouped in runs. The waveforms of the PMT and SiPM signals of 
each event in a run were stored in corresponding ROOT files for offline data analysis. The typical 
number of events for runs with the muon or hadron beam was about 5000, while for the electron beam, 
due to lower intensity, the typical number was about 600.\\

During some specific runs, a 3.7~cm thick stainless steel plate was additionally placed in front of 
the LSD module. This allowed to study the LSD response to hadronic or electromagnetic (pre-)showers 
in case where a hadron or electron traversed the LSD module, and to study in general the 
response of the LSD for higher photon yields.\\

During data taking, the output signal of the WOM with PMT1 was observed to be much smaller 
than the output signal of the WOM with PMT2. Although this was traced down to insufficient 
optical coupling between the WOM light guide and PMT1, the problem could not be solved during 
the test-beam data taking. Therfore, only waveforms for the WOM with PMT2 and the WOM with 
the SiPM array are further considered in the offline analysis.\\

\section{Data Analysis and Results}
\label{sec:dataanalysis}

\subsection{Waveform analysis and signal definition}
\label{subsec:waveform}
\begin{figure}
     \centering
     \subfigure[]{
          \label{fig:Waveforms_PMT2}
          \includegraphics[scale=0.35]{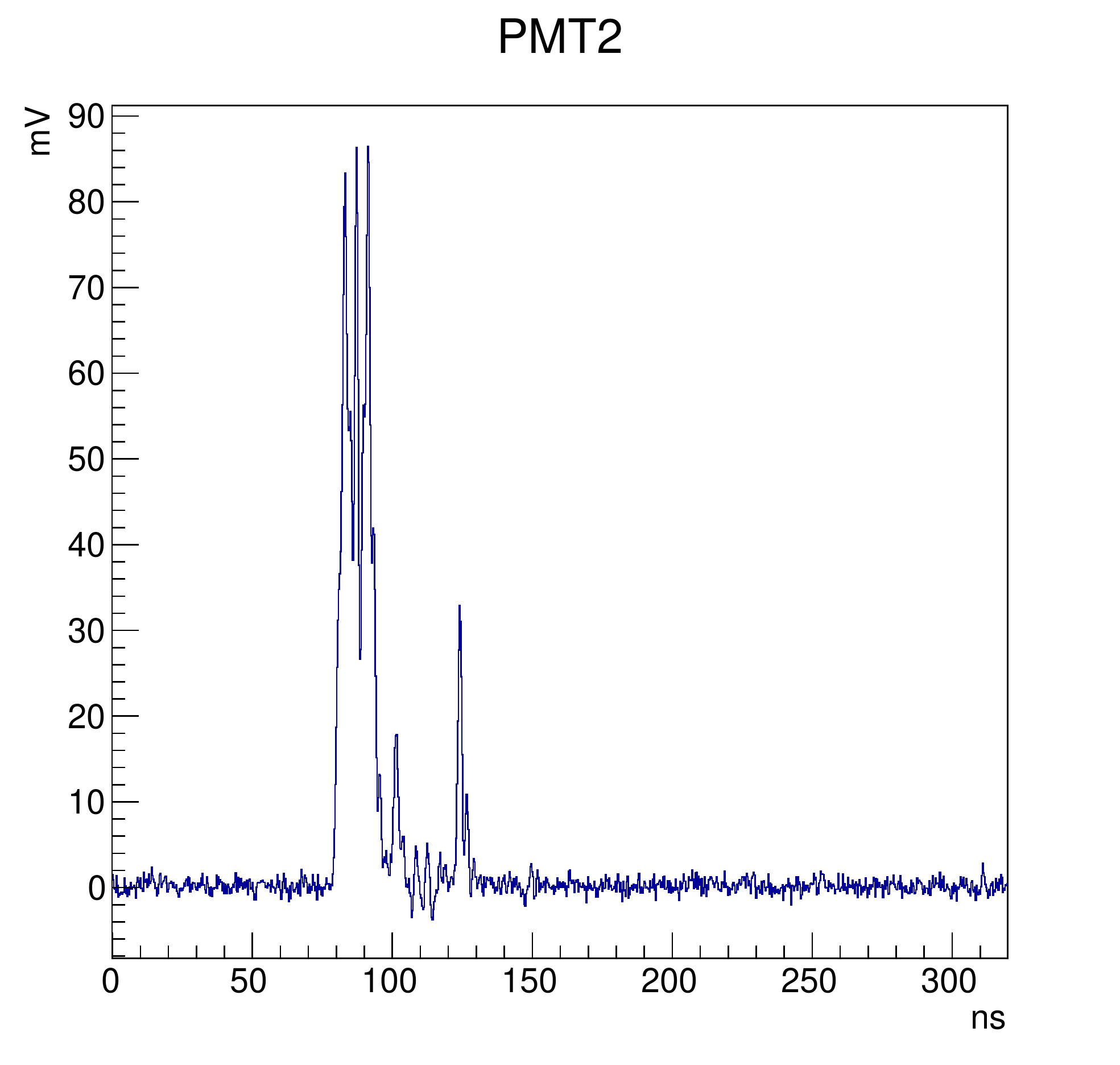}}
     \subfigure[]{
          \label{fig:Waveforms_SiPMs}
          \includegraphics[scale=0.35]{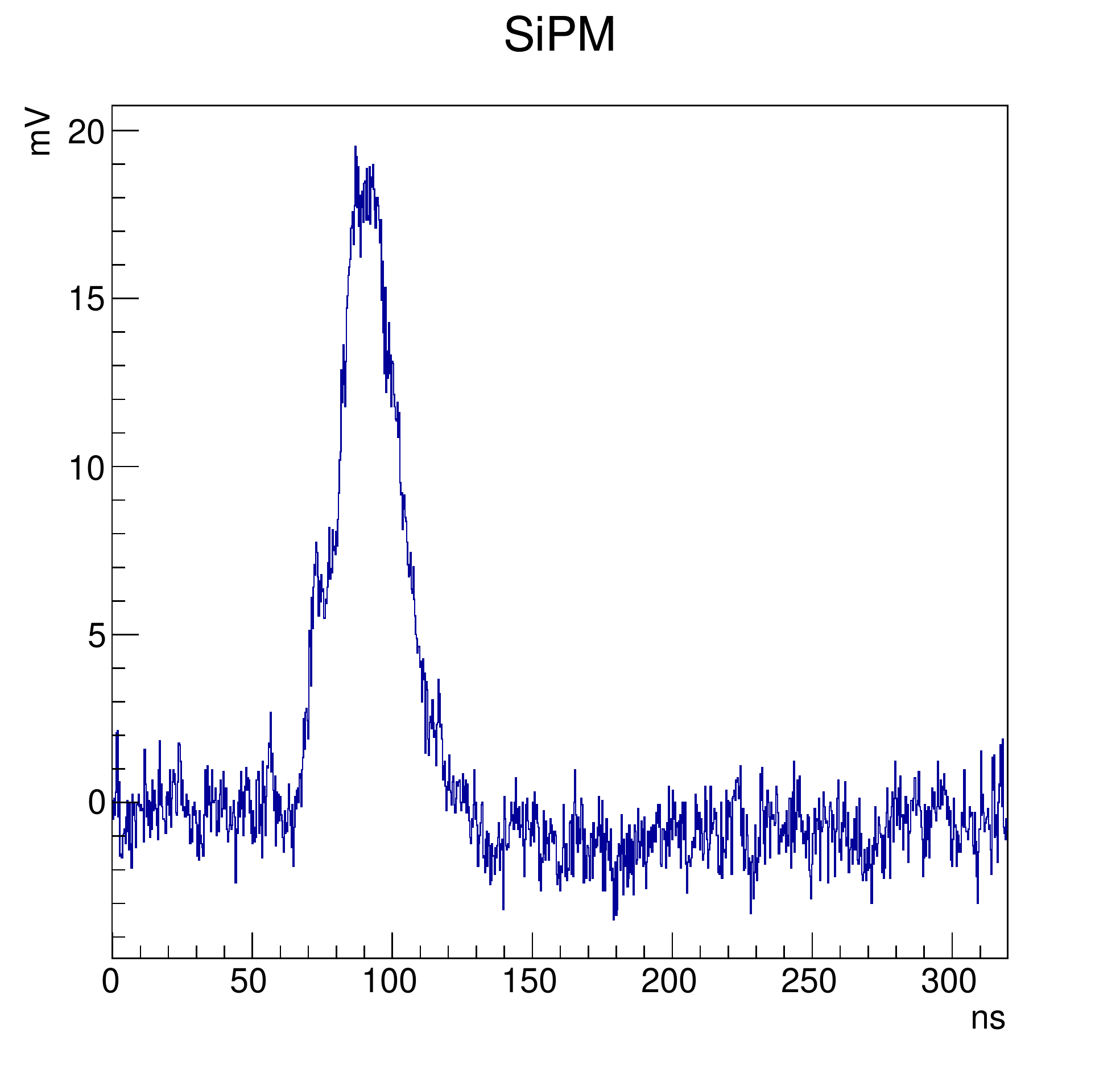}}
     \subfigure[]{
          \label{fig:Waveforms_SiPMs_SUM}
          \includegraphics[scale=0.35]{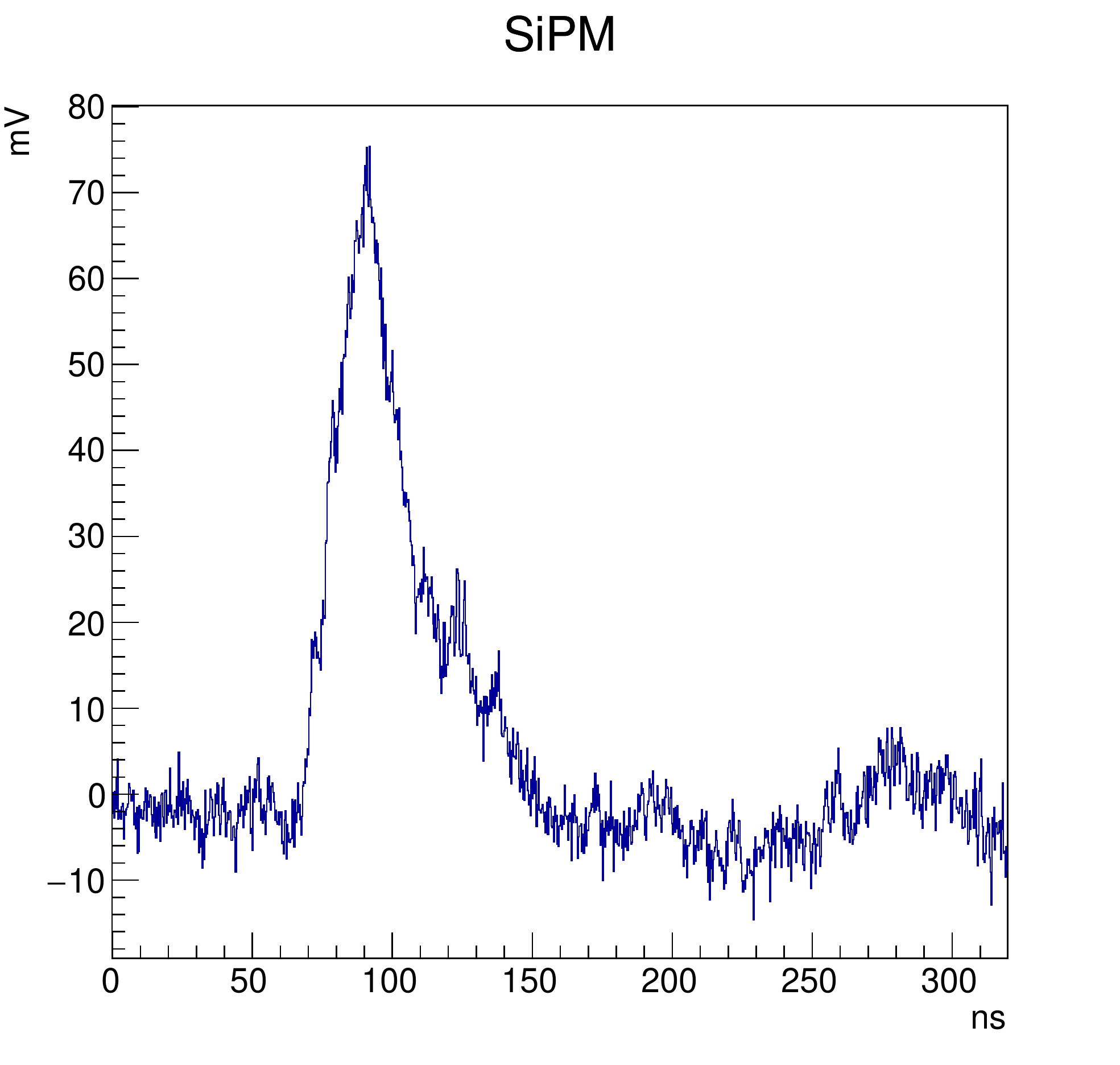}}
  \caption{Examples of a waveform for a single event from the hadron beam pointing to 
position 0 with an incident angle of $0^{\circ}$: for (a) the WOM with PMT2, 
(b) the WOM with SiPMs for one individual SiPM, and (c) the sum signal of all SiPMs.}
\label{fig:Waveforms_PMT2_SiPMs}
\end{figure}

\begin{figure}
     \centering
          \includegraphics[scale=0.35]{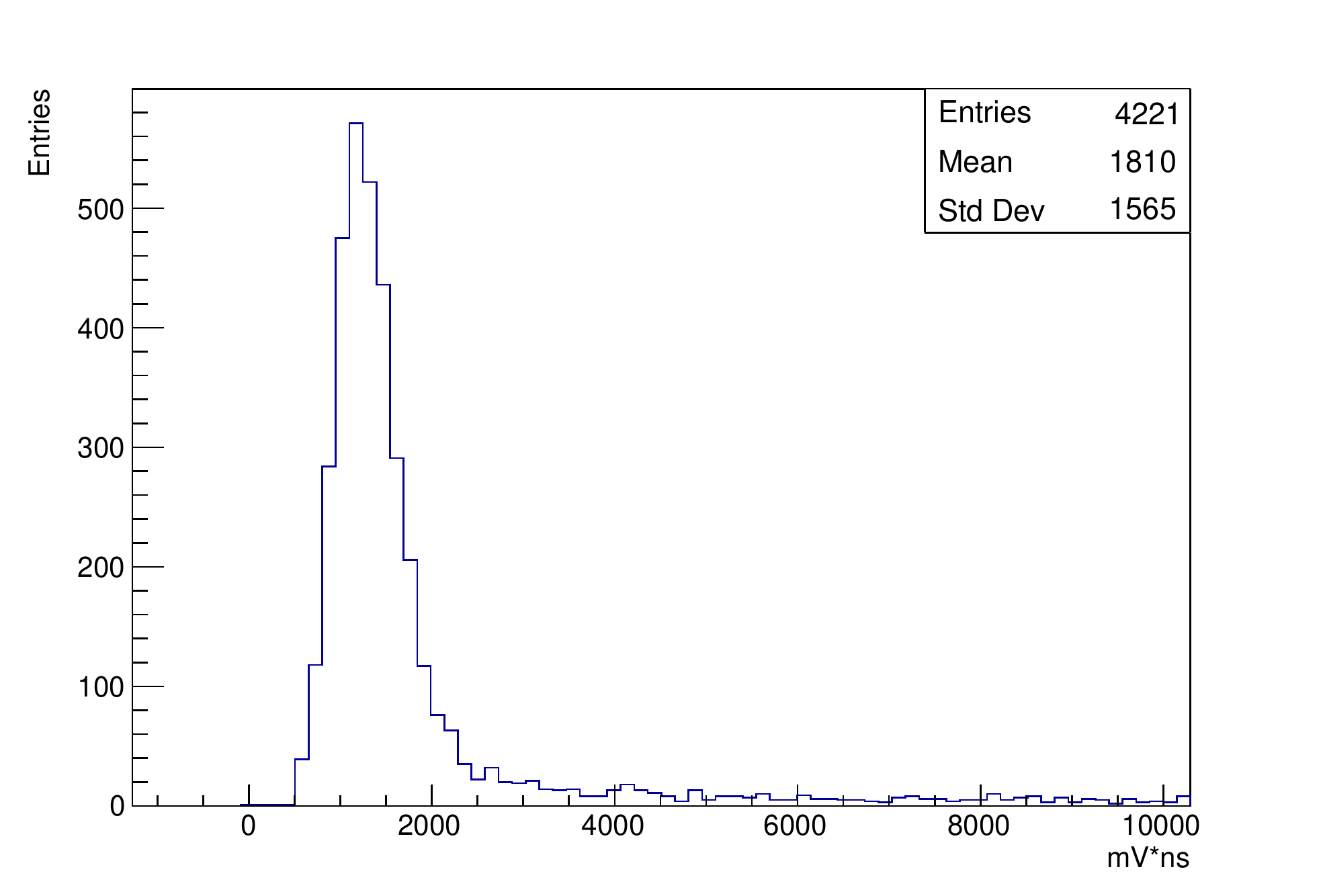}
  \caption{
Time-integrated spectrum for the WOM with PMT2 obtained for events 
from the hadron beam pointing to position 0 with an incident angle of $0^{\circ}$.}
  \label{fig:TimeIntegrated_Distribution_PMT}
\end{figure}

Fig.~\ref{fig:Waveforms_PMT2} shows a typical waveform of a signal of PMT2,
for an event in which the trigger scintillators have fired.
Fig.~\ref{fig:Waveforms_SiPMs} shows a typical waveform of an individual SiPM channel,
while Fig.~\ref{fig:Waveforms_SiPMs_SUM} shows for the same event the sum signal 
(of all eight SiPMs).
For the PMTs, the waveforms are then integrated within a 70~ns time window starting 5~ns before 
the signal (see Fig.~\ref{fig:TimeIntegrated_Distribution_PMT}). The integral is converted from 
${\rm mV} \times {\rm ns}$ to the number of photoelectrons ($N_{\rm pe}$) by means of a dedicated 
laboratory calibration measurement using the LED pulser and digitizer from the CAEN educational 
toolkit~\cite{CAENEducationalToolkit}.
In case of the SiPM array, the maximum amplitude of the waveform for each individual SiPM
channel within a 70~ns time window starting 5~ns before the signal is converted to the number 
of photoelectrons ($N_{\rm pe}$) using a corresponding calibration measurement. In the subsequent 
analysis, $N_{\rm pe}$ for the whole WOM is then estimated by adding up the seven individually 
read-out SiPM channels and scaling the result by the factor 8/7 in order to account for the eigth 
SiPM. For timing measurements, the sum signal from the MUSIC board is used.

To define the arrival time of the signals for the WOM with PMT2 (SiPM array), 
the digital constant fraction technique has been applied~\cite{Delagnes}:
The maximum of the signal within a time window of 320~ns is searched for and then 
the earliest time before the maximum is taken, for which the signal has $20~\%$ ($30~\%$) 
of the maximum signal height. This threshold value for the constant fraction discrimination 
is chosen as the result of an optimization procedure that determines the minimal 
standard deviation of the arrival time as a function of the threshold. For the WOM 
with the SiPM array, only the sum channel of the SiPMs is considered if not explicitly 
stated otherwise.

It was explicitly checked that there are no events with more than one trigger signal
within a time window of 320~ns, so that the probability of two beam particles hitting
the LSD module in one event is negligible.

\subsection{Photo detection efficiency of the liquid-scintillator detector}
\label{subsec:efficiency}
\begin{figure}
     \centering
     \subfigure[]{
          \label{fig:PE_Distribution_PMT}
          \includegraphics[scale=0.35]{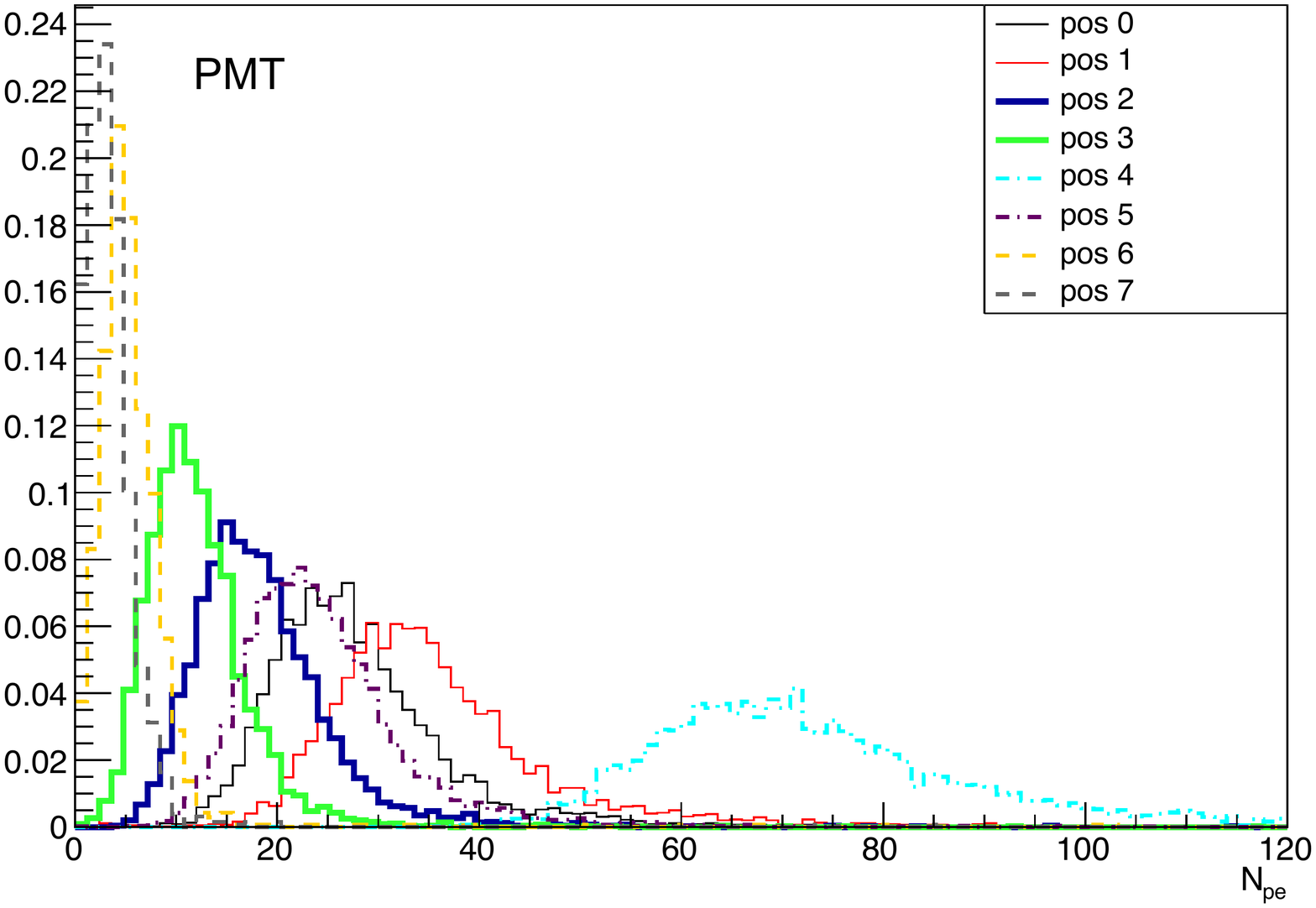}}
     \subfigure[]{
          \label{fig:PE_Distribution_SiPM}
          \includegraphics[scale=0.35]{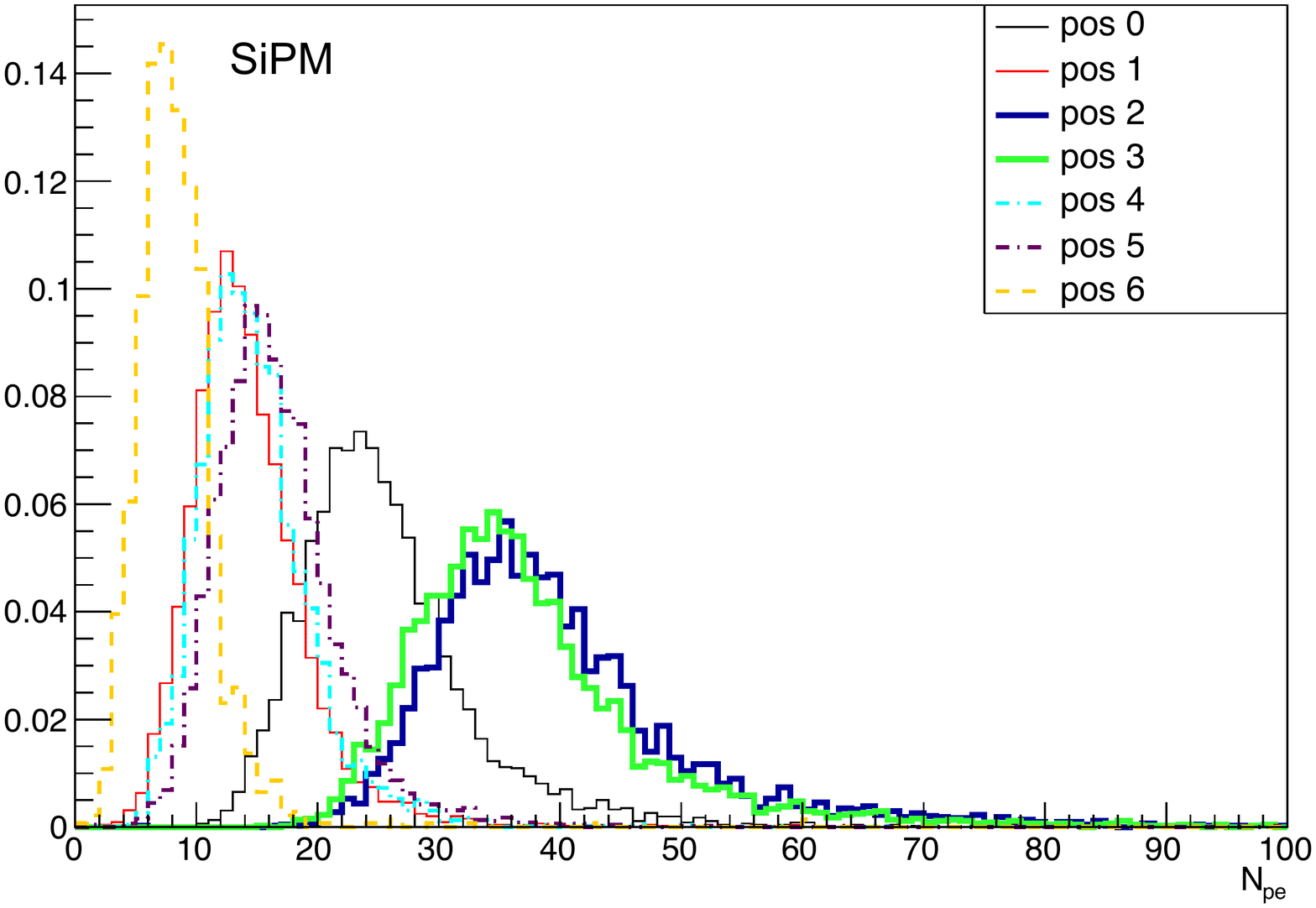}}
  \caption{
$N_{\rm pe}$ distribution for muons and different beam positions 
at an incident angle of $0^{\circ}$. (a) WOM with PMT2 and (b) WOM with SiPM array.
For the latter case, $N_{\rm pe}$ at position 7 is not shown, since it is much higher 
compared to positions 0 -- 6 because the particle beam directly hits the WOM with 
the SiPM array.}
  \label{fig:PE_Distribution_PMT_SiPM}
\end{figure}

For the different particle types, the photo detection efficiency of the LSD module looking 
only at a specific WOM is determined as a function of $N_{\rm pe}$ for different beam positions 
on the LSD module and an incident angle of $0^{\circ}$. The photo detection efficiency is 
defined as the ratio between the number of events with a WOM signal above a certain 
photoelectron threshold and the total number of triggered events. 
\begin{figure}
     \centering
     \subfigure[]{
          \label{fig:Efficiency_PMT2_2PE}
          \includegraphics[scale=0.35]{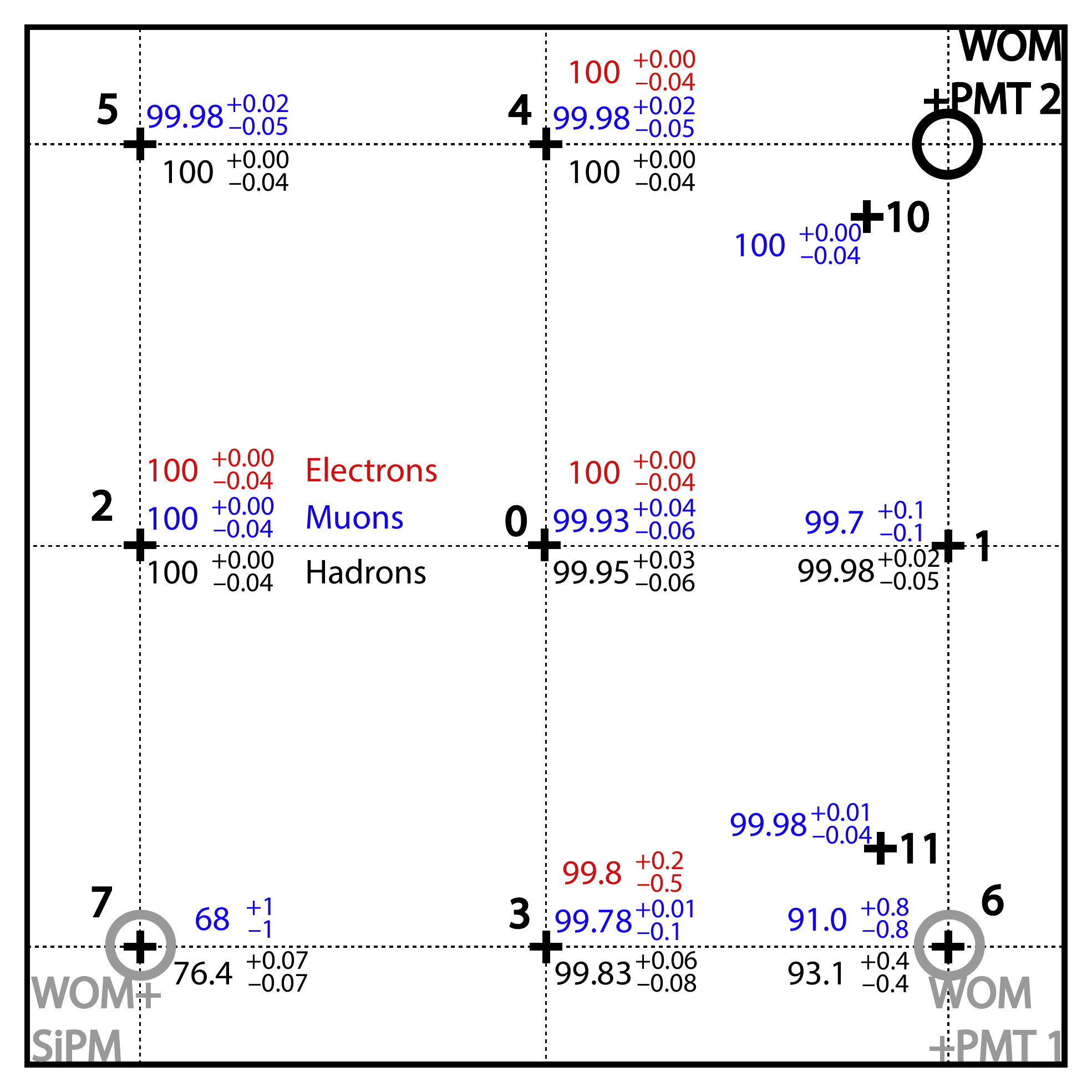}}
     \subfigure[]{
          \label{fig:Efficiency_SiPMs_2PE}
          \includegraphics[scale=0.35]{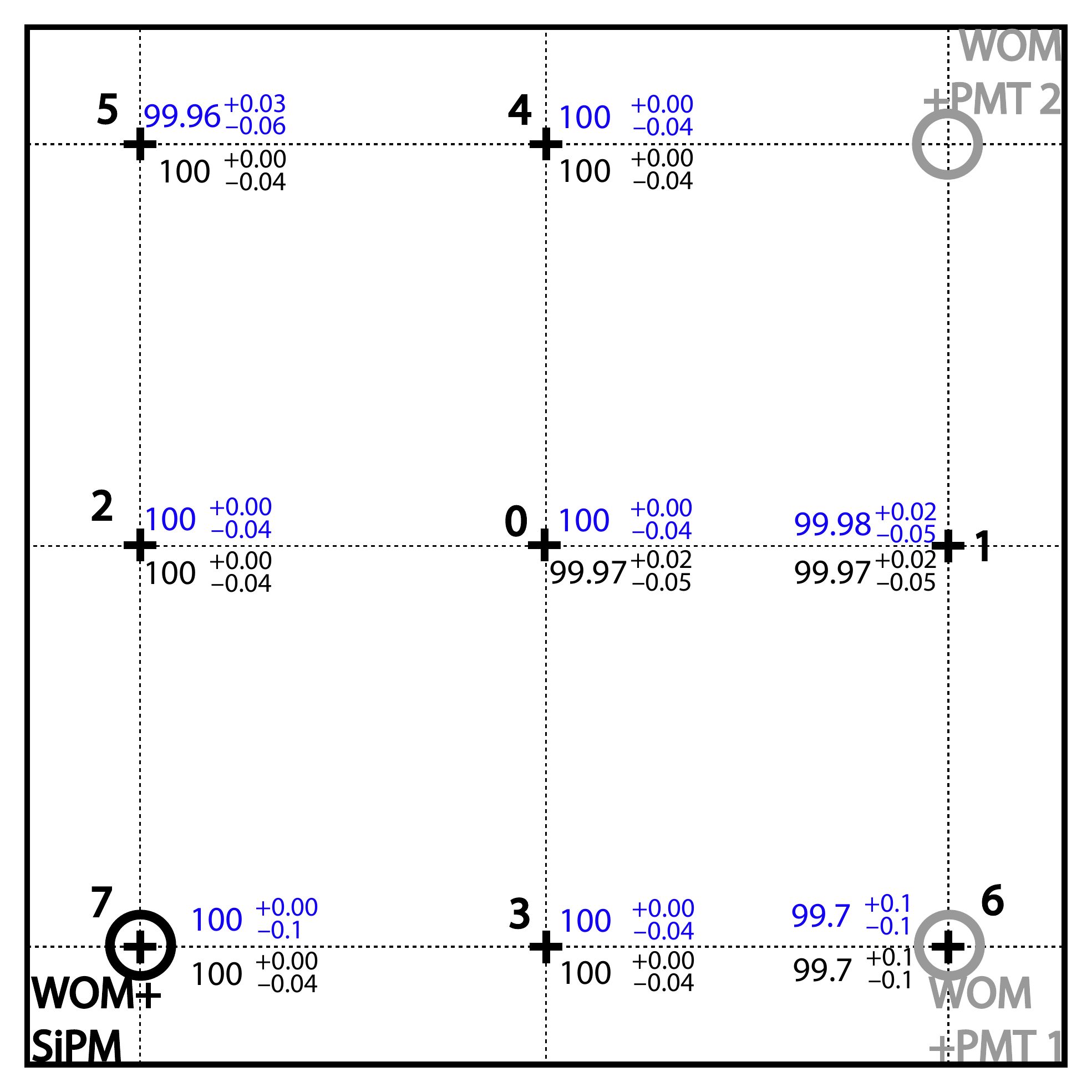}}
  \caption{Photo detection efficiency for a threshold of $N_{\rm pe} \ge 2$ for (a) the WOM with PMT2 
and (b) the WOM with SiPMs for different beam positions (indicated by black crosses) 
at an incident angle of $0^{\circ}$. Efficiency numbers are quoted from bottom to top for hadrons, muons, and for electrons. Numbers for positions 10 and 11 for the WOM with 
SiPMs are not quoted, since these later measurements were suffering from liquid-scintillator leakage 
into the PMMA vessel, lowering the efficiency .}
  \label{fig:Efficiency_PMT2_SiPMs_2PE}
\end{figure}
Figs.~\ref{fig:Efficiency_PMT2_2PE} and~\ref{fig:Efficiency_SiPMs_2PE} show the photo detection efficiency 
for a threshold of $N_{\rm pe} \ge 2$ for the WOM with PMT2 respectively for the one with the SiPM array for 
different beam positions on the LSD module (indicated by black crosses) and an incident angle of $0^{\circ}$. 
The quoted uncertainties are estimated for $68~\%$ confidence level using the Clopper-Pearson method. 
For the WOM with PMT2, the photo detection efficiency is presented for all positions and is better than
$99.7\%$ for any particle type and all beam positions, except those where the beam directly hits another 
WOM (positions 6 and 7). 
For these positions, the photo detection efficiency of the WOM with PMT2 is significantly reduced, 
because a large part of the scintillation light is generated inside the other WOM and absorbed with 
very large probability right away in its wavelength-shifting paint. 
A similar effect is seen for the WOM with the SiPM array: 
when the beam hits position 6 where another WOM is located, the photo detection efficiency is smaller 
than at other positions, although the effect is less pronounced compared to the photo detection 
efficiency of the WOM with PMT2. On the other hand, when the beam directly hits the studied WOM, 
the photo detection efficiency is very large, see position 7 in Fig.~\ref{fig:Efficiency_SiPMs_2PE}.
\begin{figure}
     \centering
     \subfigure[]{
          \label{fig:Efficiency_PMT2_5PE}
          \includegraphics[scale=0.35]{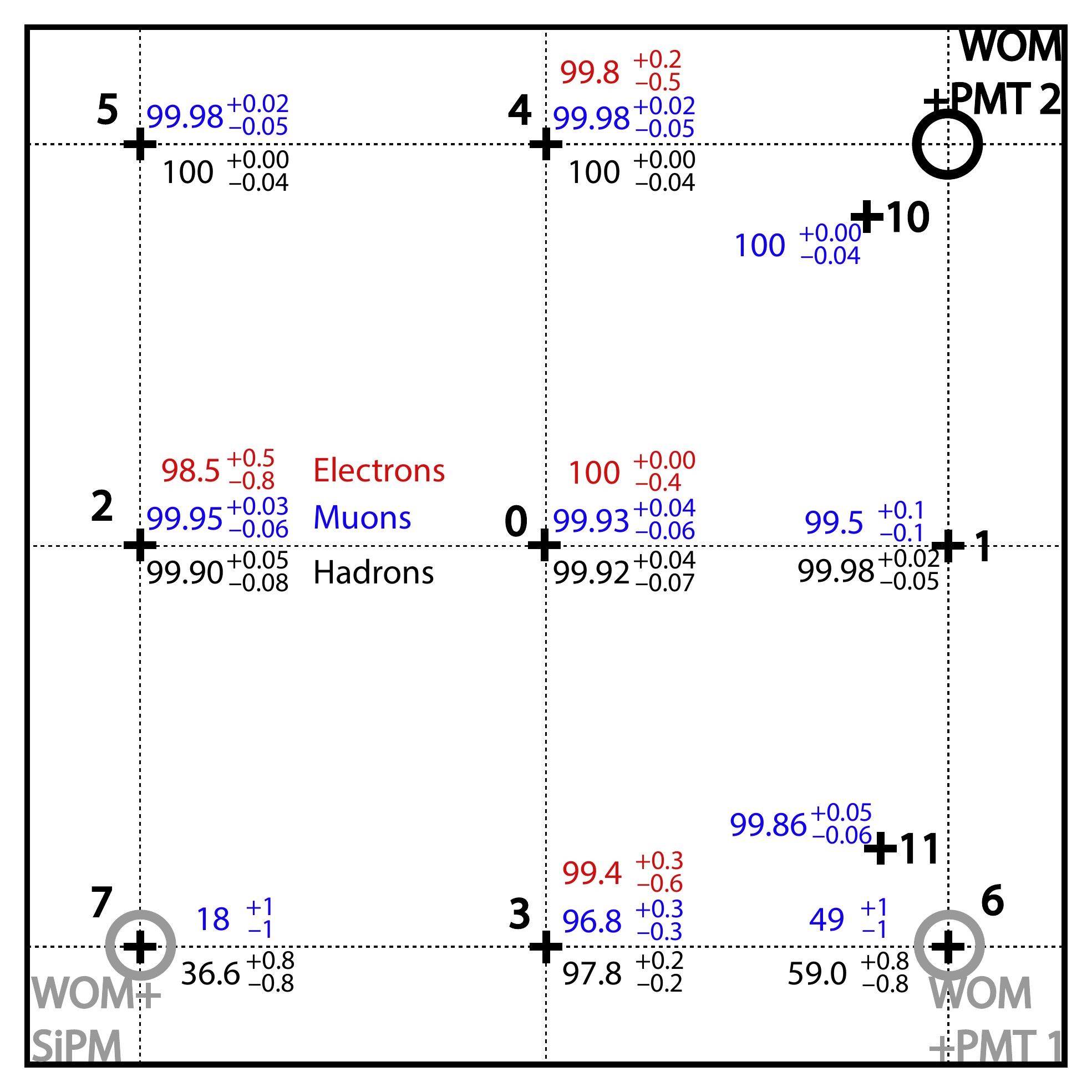}}
     \subfigure[]{
          \label{fig:Efficiency_SiPMs_5PE}
          \includegraphics[scale=0.35]{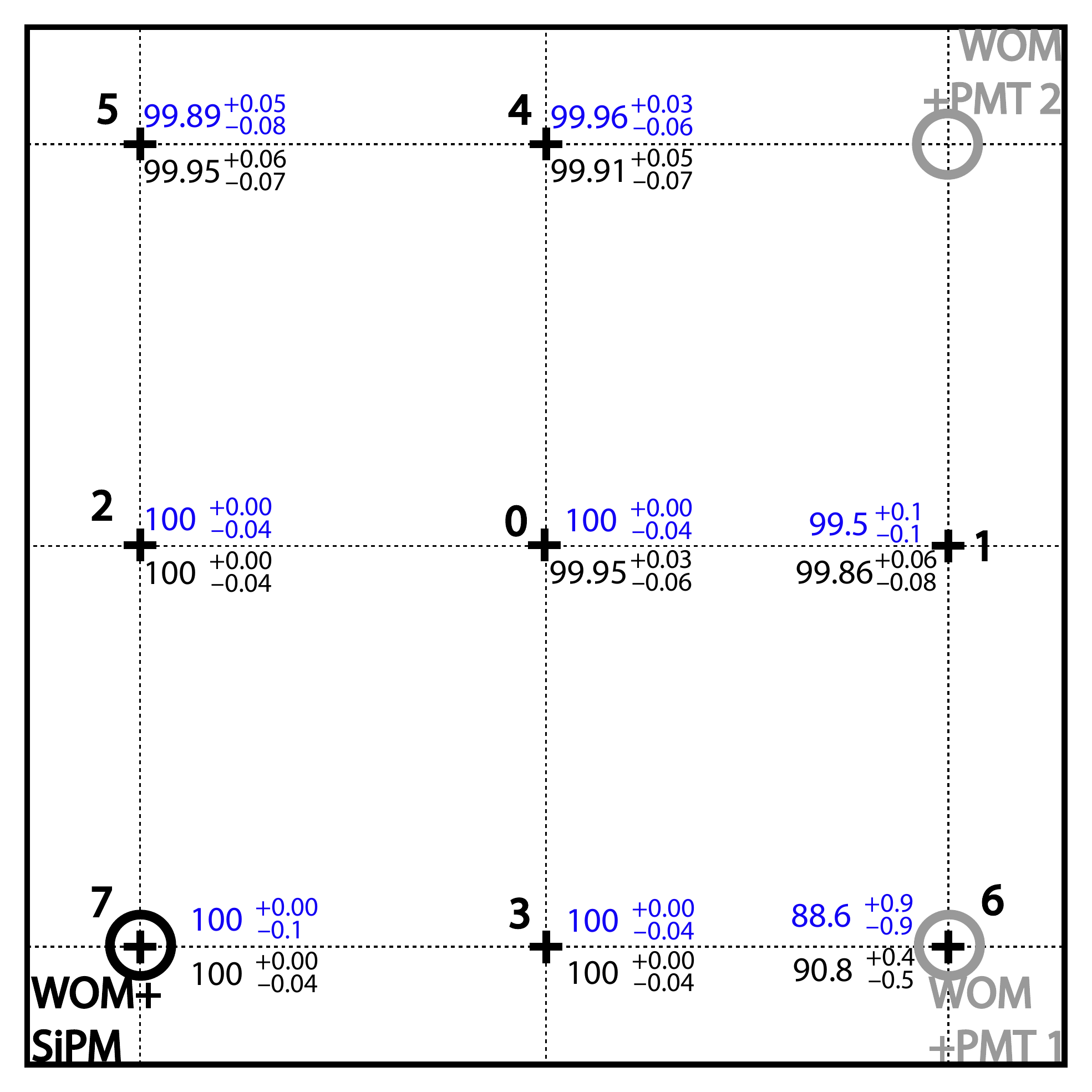}}
  \caption{Photo detection efficiency for a threshold of $N_{\rm pe} \ge 5$ for (a) the WOM with 
PMT2 and (b) the WOM with SiPMs for different beam positions (indicated by black crosses) at an 
incident angle of $0^{\circ}$. Efficiency numbers are quoted from bottom to top for hadrons, 
muons, and for electrons.}
  \label{fig:Efficiency_PMT2_SiPMs_5PE}
\end{figure}
Figs.~\ref{fig:Efficiency_PMT2_5PE} and~\ref{fig:Efficiency_SiPMs_5PE} show the photo detection efficiencies 
for a threshold of $N_{\rm pe} \ge 5$. For the WOM with PMT2 at position 3, the efficiency drops to $98~\%$, 
since the average number of photoelectrons for this position is only O(10). This is visualized in 
Figs.~\ref{fig:PE_Distribution_PMT} and~\ref{fig:PE_Distribution_SiPM} showing the $N_{\rm pe}$ distribution 
for the WOM with PMT2 and the WOM with the SiPM array at different beam positions. 
As expected, the $N_{\rm pe}$ yield decreases with increasing distance between the particle track and the 
studied WOM, and when the beam hits a different WOM, the $N_{\rm pe}$ yield at the studied WOM is particularly 
low, as discussed above.\\
Several factors may be optimised to improve the detected light yield:
\begin{itemize}
\item The PPO concentration of 1.5~g/l does not give the maximum light yield and could be 
      increased~\cite{Djurcic:2015vqa} with the positive side effect that also the time 
      resolution of the liquid scintillator would improve~\cite{MarrodanUndagoitia:2009kq}.      
\item The absorption length of the liquid scintillator reduces the light yield. It could be increased 
      by a dedicated $Al_{2}O_{3}$ purification procedure~\cite{Djurcic:2015vqa,Benziger:2007aq}.
\item Equipping all of the inner walls of the LSD module with a reflective foil or coating.
\item The photo sensor coverage of the light guide exit can be optimised.
      In particular, the fact that the PMT does not cover the whole light guide exit reduces its
      collected light by a significant amount. The observed photoelectron yield for the WOM with PMT2 is 
      smaller on average than that of the WOM with SiPMs (see Figs.~\ref{fig:PE_Distribution_PMT} 
      and~\ref{fig:PE_Distribution_SiPM}), despite the larger coverage of the light guide exit  
      by PMT2. Indeed, the individual SiPM photoelectron yields show that a SiPM placed on the outer 
      ring of the SiPM array detects about a factor of two more photons than the SiPM in the centre
      of the array. Compared to the PMT case, the detection efficiency is further increased for 
      the SiPMs case by their higher photo detection efficiency in particular in the wavelength
      range of the WLS emission spectrum (see Fig.~\ref{fig:EmissionAbsorptionSensitivities}).
\item After the test-beam measurements, it has been found with a setup at the 
      University of Mainz~\cite{WOMefficiencysetup} that the WOM with the 
      light guide has a photo detection efficiency that is about one order of magnitude lower 
      than a WOM without a light guide~\cite{MainzIceCubeTest}. 
      Possible reasons are: I. significant light losses in the light guide 
      due to its geometrical shape, which is supported by GEANT4-based simulations~\cite{GEANT4_1,GEANT4_2};
      II. the quality of the gluing between the WOM tube and the light guide.\\
      Problem II could be circumvented if the WOM tube and the light guide are made out of one piece.
      Problems I and II could be both avoided by not relying on a light guide, but 
      placing e.g. SiPMs on a ring-like structure and coupling them directly to the end of the WOM tube. 
\item Furthermore, the PMMA used for the WOM vessels could be optimised, as after the test beam measurements, 
      the employed material was found to not provide a maximal transmission coefficient in the PPO emission 
      wavelength range. 
\end{itemize}

\subsection{Time Resolution}
\label{subsec:timeresolution}
\begin{figure}
  \begin{center}
             \includegraphics[scale=0.50]{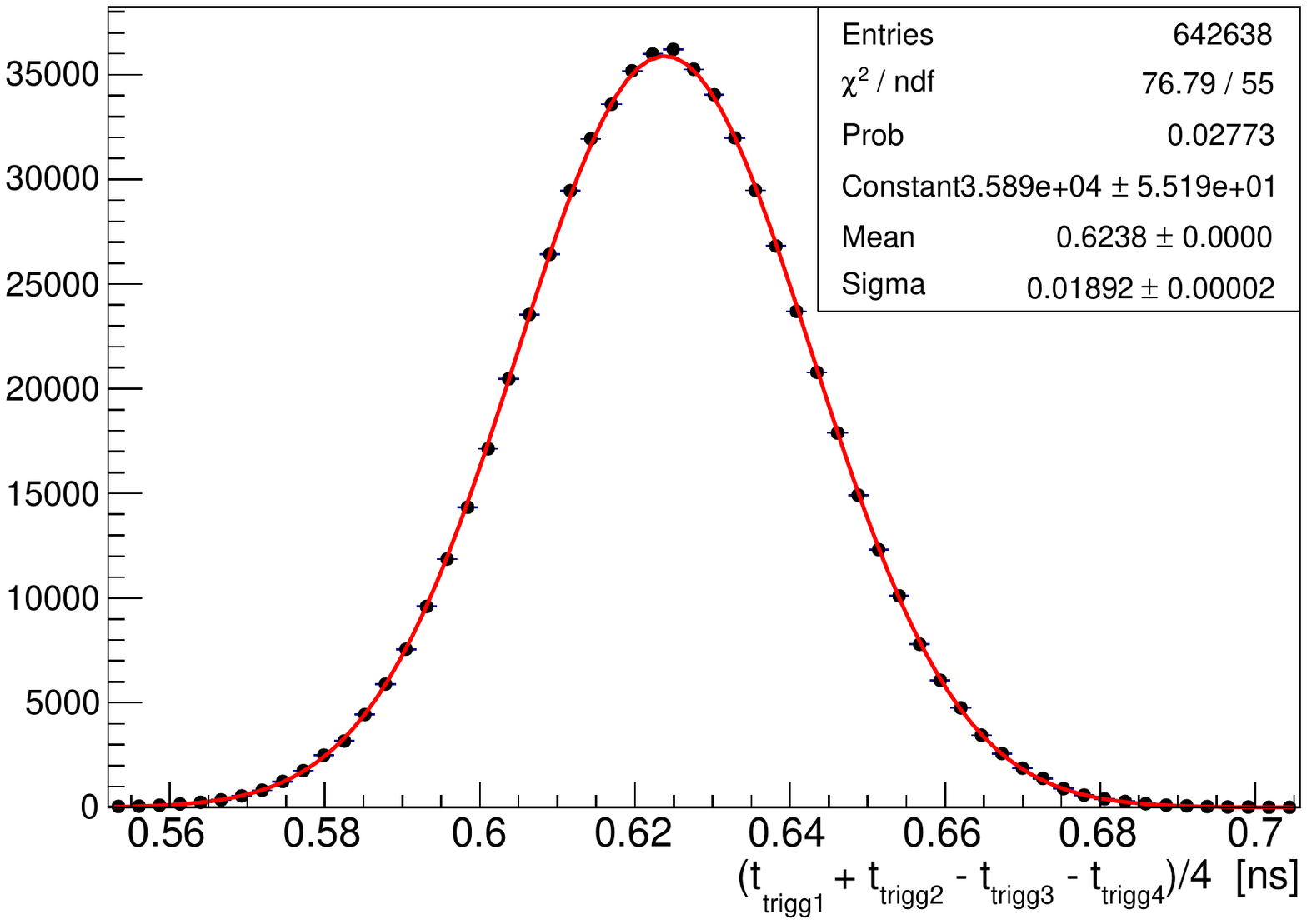}
  \end{center}
  \caption{Average time distribution of the four PMTs used for the two trigger
  scintillators $(t_{trigg1}+t_{trigg2}-t_{trigg3}-t_{trigg4})/4$. A Gaussian
  fit to the distribution gives a standard deviation of 18.9 ps.}
  \label{fig:TriggerTimeResolution}
\end{figure}

\begin{figure}
     \centering
     \subfigure[]{
          \label{fig:DeltaT_PMT2}
          \includegraphics[scale=0.35]{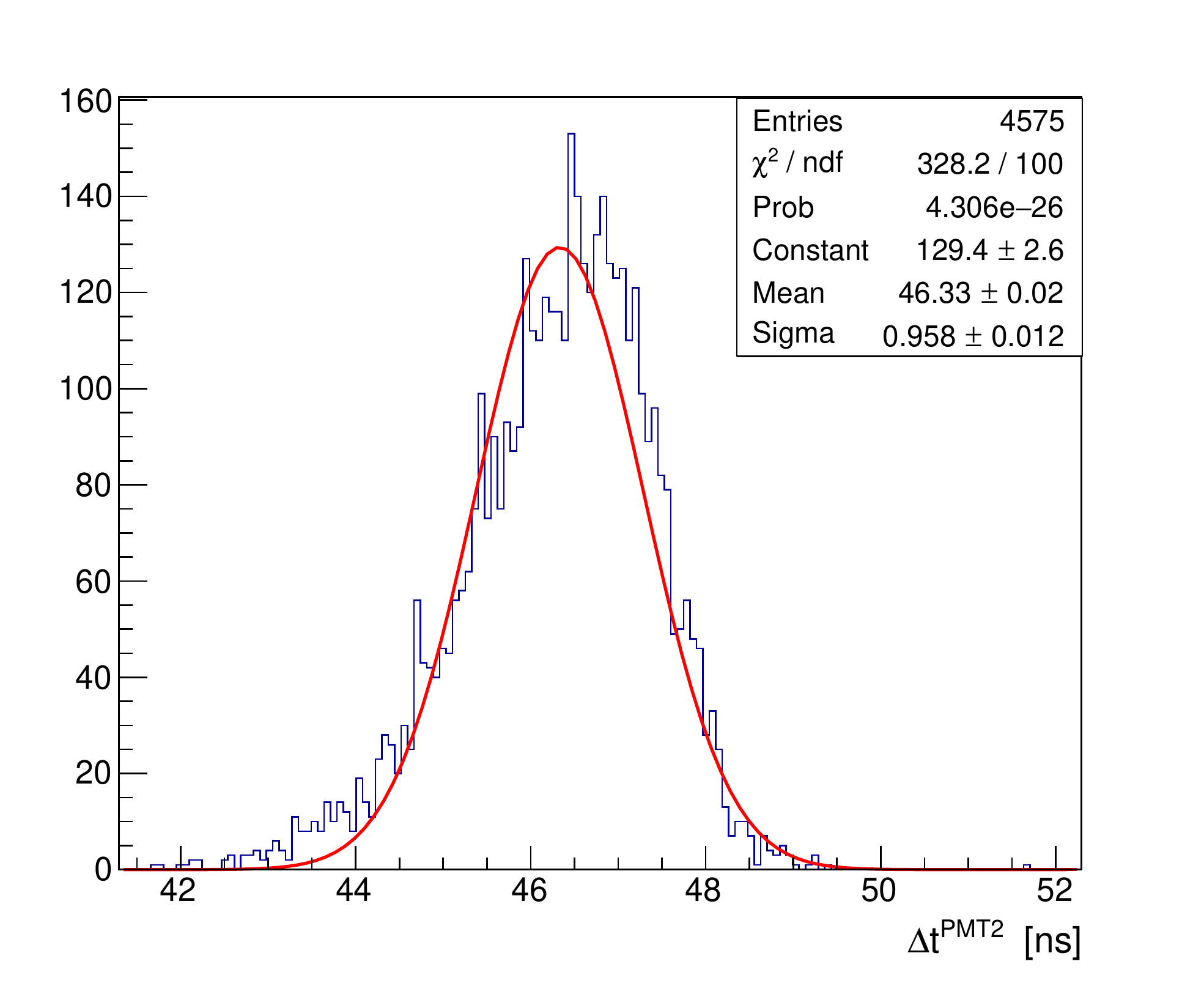}}
     \subfigure[]{
          \label{fig:DeltaT_SiPMs}
          \includegraphics[scale=0.35]{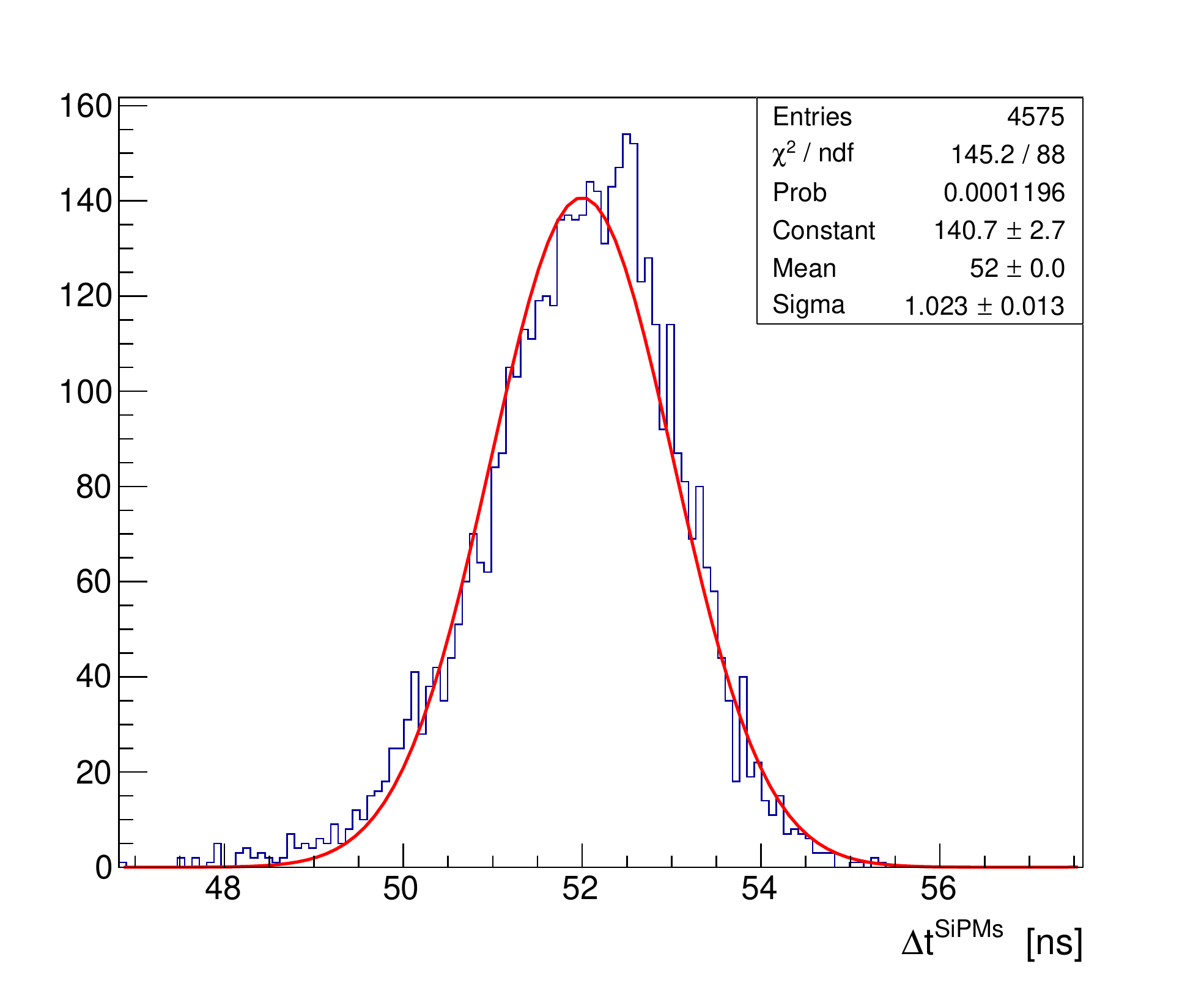}}
  \caption{$\Delta t$ distributions for muons at an incident angle $0^{\circ}$ 
and beam position 0: (a) WOM with PMT2, $\Delta t^{PMT2}$, (b) WOM with SiPMs, 
$\Delta t^{SiPMs}$.}
  \label{fig:DeltaT_PMT2_SiPMs}
\end{figure}

\begin{figure}
  \centering
  \includegraphics[scale=0.50]{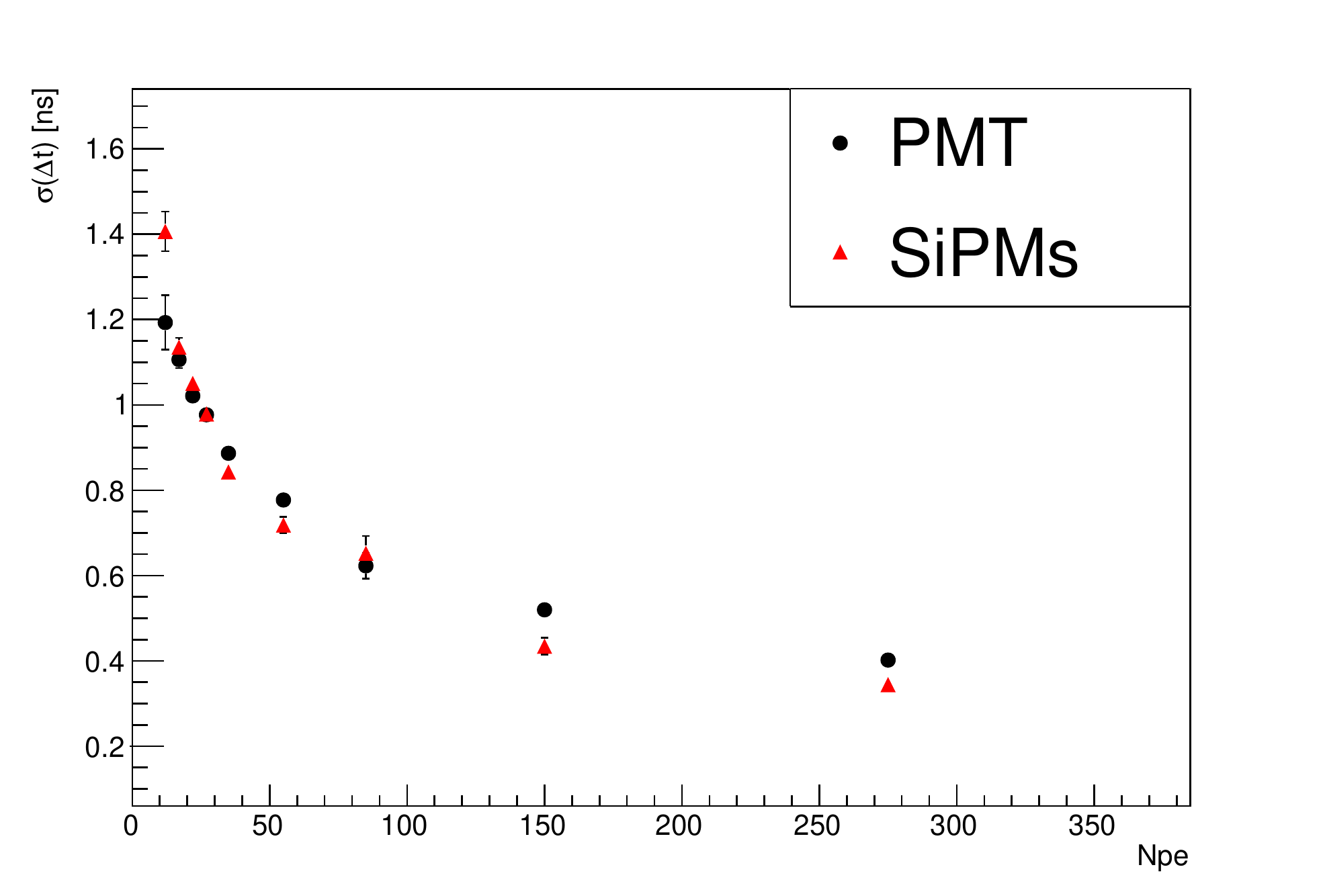}
  \caption{$\sigma(\Delta t)$ as a function of the number of photoelectrons for
different beam positions at an incident angle $0^{\circ}$ for the WOM with PMT2 
and the WOM with SiPMs.}
  \label{fig:DeltaT_PMT2_SiPMs_PE}
\end{figure}

To measure the time resolution of the LSD module, the difference 
\begin{equation}
\Delta t= t_{trigg}-t_{WOM}
\end{equation}
between the average time of the trigger scintillators 
\begin{equation}
t_{trigg}=(t_{trigg1}+t_{trigg2}+t_{trigg3}+t_{trigg4})/4
\end{equation} 
and the WOM signal from PMT2 respectively the SiPM array is determined ($t_{WOM}$). 
Fig.~\ref{fig:TriggerTimeResolution} shows the time distribution of 
$(t_{trigg1}+t_{trigg2}-t_{trigg3}-t_{trigg4})/4$ for triggered events, which demonstrates 
the time resolution of the trigger system. The standard deviation $\sigma$ of a 
Gaussian fit to this distribution is found to be $18.9~{\rm ps}$. Since the time 
resolution of the LSD module is much larger due to the fast component of the 
scintillation decay time of 6~ns~\cite{MarrodanUndagoitia:2009kq}, 
the effect of $t_{trigg}$ on the $\Delta t$ resolution is negligible.
Figs.~\ref{fig:DeltaT_PMT2} and~\ref{fig:DeltaT_SiPMs} show typical $\Delta t$ spectra 
for the WOM with PMT2 and for the WOM with the SiPM array respectively. The time resolution 
is then given by the standard deviation $\sigma$ of a Gaussian fit to this distribution. 
For the SiPM array, a dependence of the signal time on its amplitude was observed.
The average time-walk effect could be parameterized by a 4th-order polynomial, 
fit to the two-dimensional $\Delta t^{SiPMs}$ vs. $N_{\rm pe}$ distribution.
Fig.~\ref{fig:DeltaT_PMT2_SiPMs_PE} shows the time resolution as a function of the number 
of photoelectrons for the WOM with PMT2 respectively with the SiPM array, after application 
the time-walk correction. The time resolution improves with increasing number of photoelectrons 
and is better than 1~ns for more than $N_{\rm pe} > O(25)$ for both WOMs.\\
 
\begin{figure}
  \begin{center}
             \includegraphics[scale=0.50]{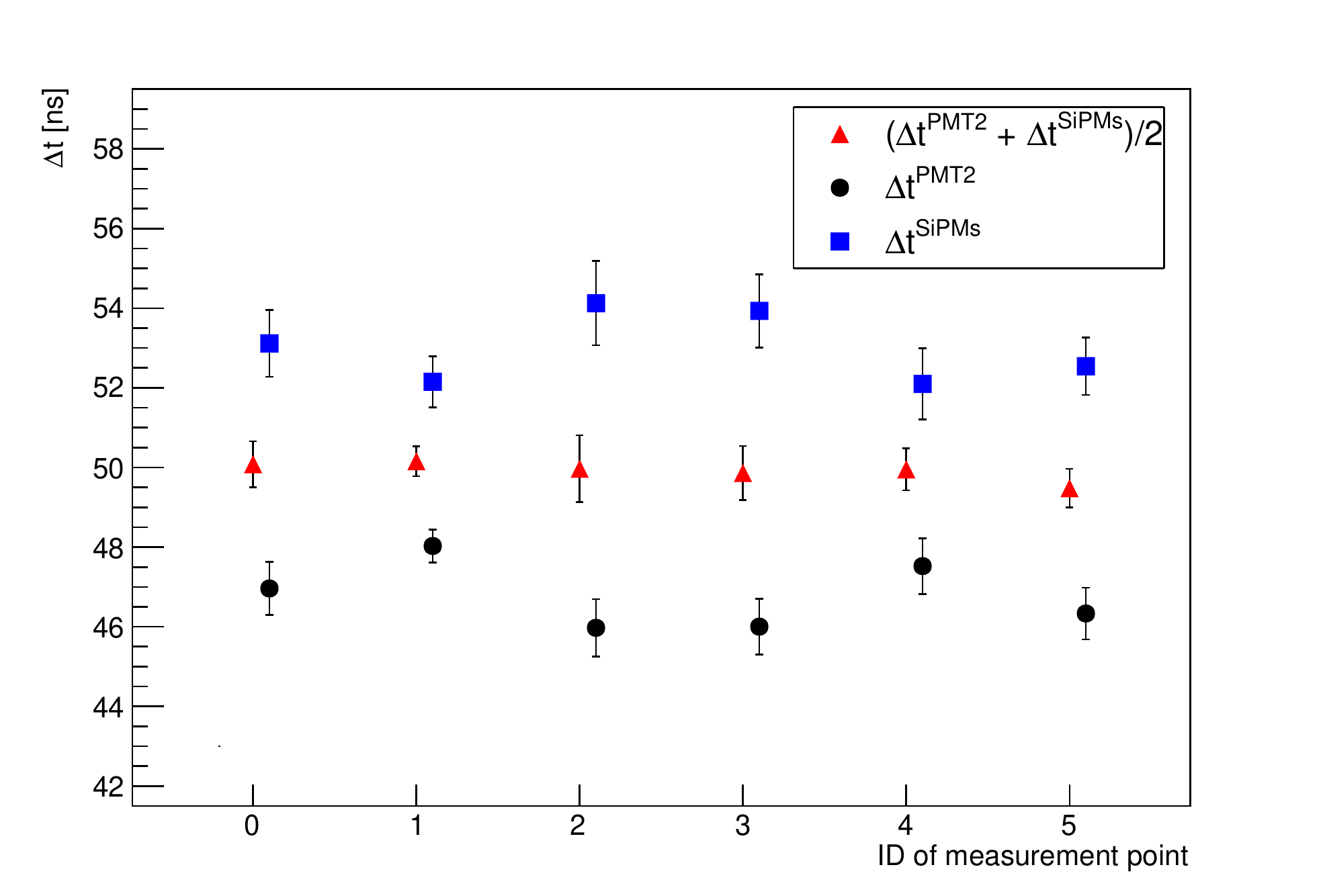}
  \end{center}
  \caption{Mean value of the $\Delta t^{PMT2}$, $\Delta t^{SiPMs}$, and
           $(\Delta t^{PMT2}+\Delta t^{SiPMs})/2$ distributions for various 
           beam positions on the LSD module at an incident angle of $0^{\circ}$.
           The error bars represent the standard deviation (i.e. the time resolution)
           of the corresponding distribution.}
  \label{fig:MeanAndStandardDeviationTimeResolution}
\end{figure}

In a real experiment, the position where a particle hits the LSD module might not be known from 
other sub-detector information. This will add an additional time uncertainty due to the geometrical 
extension of the LSD module leading to different light-traveling paths inside the liquid scintillator. 
By using the average time of both WOMs, the time response of the LSD module becomes less sensitive 
to this geometry effect. This can be seen in Fig.~\ref{fig:MeanAndStandardDeviationTimeResolution}, 
showing the mean of the individual time distributions for the two WOMs, 
$\Delta t^{PMT2}$ and $\Delta t^{SiPMs}$, as well as of their average, 
$(\Delta t^{PMT2}+\Delta t^{SiPMs})/2$, for various beam positions on the LSD module 
at an incident angle of $0^{\circ}$. With the low detected photon yield of the current design, 
the LSD module has a time resolution of about 1~ns (represented by the error bars in Fig.~\ref{fig:MeanAndStandardDeviationTimeResolution}), 
independent from the beam position.
As expected, differences between the mean values of the average time distribution, 
$(\Delta t^{PMT2}+\Delta t^{SiPMs})/2$, 
for various beam positions are much smaller than those of the individual times. 
The mean value of $(\Delta t^{PMT2}+\Delta t^{SiPMs})/2$ decreases when moving from position 0, 
(located on the diagonal between the WOM with SiPMs and the WOM with PMT2) to positions 1, 2, 3, 
and 4 (all at the same distance to that diagonal) to position 5 (the largest 
distance to the diagonal) as expected. This shift induced by the geometrical extension of 
the LSD module and the WOM positions inside the LSD module is smaller than 1 ns and 
therefore only affects the time resolution of the LSD significantly, if the photon yield is 
substantially increased or the geometrical dimensions of the LSD module are enlarged. 
This timing uncertainty can be reduced by using more than two WOMs in the LSD module.

\subsection{Detector response to a well-defined energy deposition}
\label{subsec:detectorresponse}

\begin{figure}
  \begin{center}
             \includegraphics[scale=0.50]{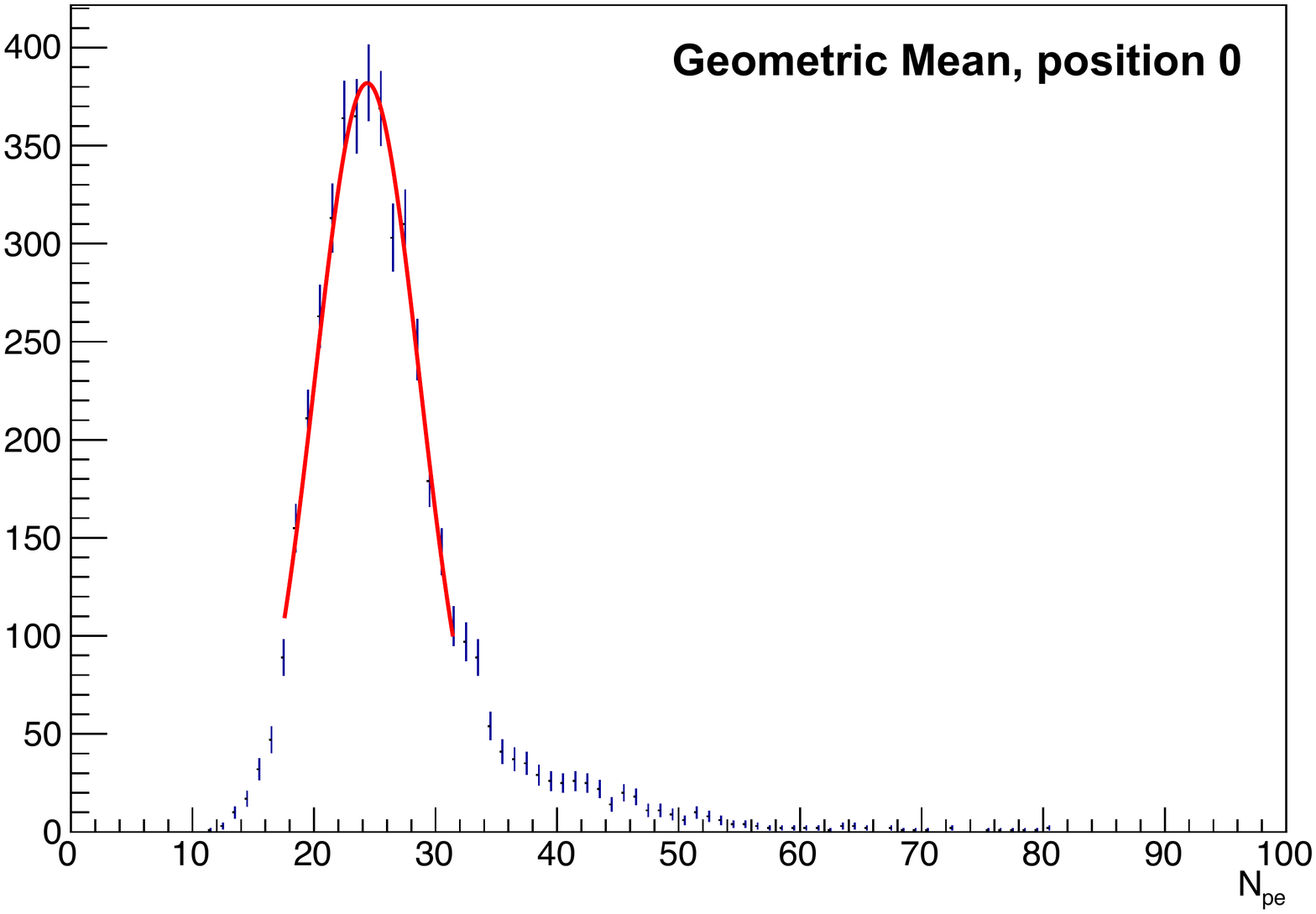}
  \end{center}
  \caption{The distribution of the geometric mean for muons hitting
           the LSD at beam position 0. A Gaussian fit is applied to the core part of the
           distribution to obtain the position of the maximum and FWHM of the distribution.}
  \label{fig:gaus_fit_example}
\end{figure}

\begin{figure}
  \begin{center}
             \includegraphics[scale=0.50]{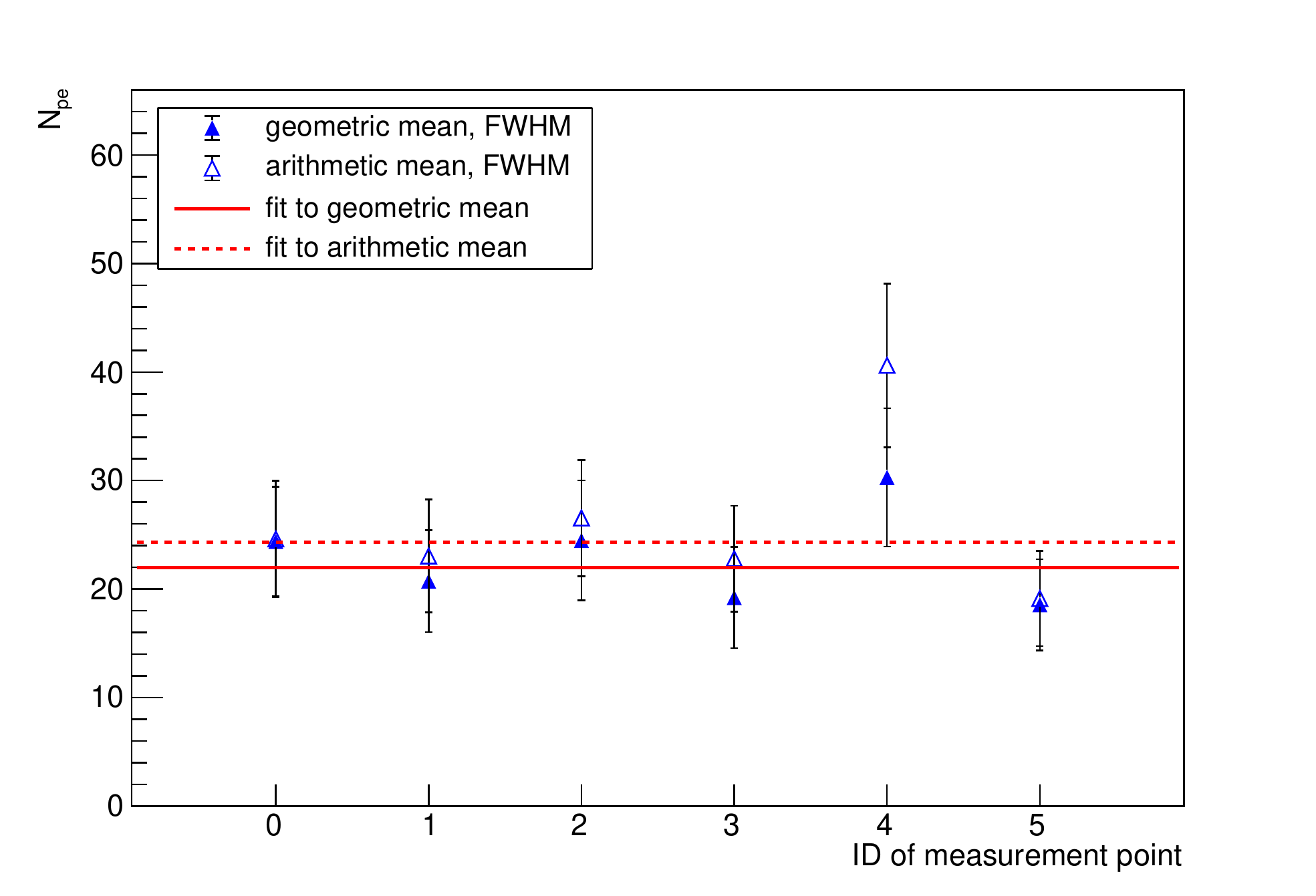}
  \end{center}
  \caption{Position of the maximum and FWHM (represented by error bars) for the distribution of
geometric and arithmetic mean of the number of photo electrons calculated from the two WOMs, 
requiring at least two photo electrons. Measurements of muons for different beam positions on 
the detector module with an incident angle of $0^{\circ}$. The average value for all beam positions 
is shown by the horizontal line.
}
  \label{fig:MeanAndGeometricalMeanVsPosition}
\end{figure}

Another important question is the homogeneity of the LSD response with respect to different beam positions. 
Two different combinations of the signals from PMT2 and the SiPMs are considered in this paper: 
   \begin{equation}
   {\rm the~arithmetic~mean}=\frac{N_{pe}^{PMT2}+N_{pe}^{SiPMs}}{2}
   \end{equation}
and 
   \begin{equation}
   {\rm the~geometric~mean}=\sqrt{N_{pe}^{PMT2} \cdot N_{pe}^{SiPMs}} ,
   \end{equation}
requiring $N_{\rm pe} \ge 2$.
The latter one is expected to provide a homogeneous response in case that the detected 
light yield drops exponentially as a function of distance between light source to WOM tube.
Fig.~\ref{fig:gaus_fit_example} shows an example of the distribution of the geometric mean 
obtained for muons hitting the LSD module at position 0. 
The value of the geometric mean at the maximum of the distribution obtained from a Gaussian 
fit to the core part of the distribution serves as an estimator for the detector response. 
For the resolution, full-width-half-maximum (FWHM) of this Gaussian function is quoted.
Fig.~\ref{fig:MeanAndGeometricalMeanVsPosition} shows the position of the maximum and
the corresponding FWHM for the distribution of the geometric mean and of the arithmetic mean
at different beam positions for muons hitting the LSD module with an incident angle of $0^{\circ}$.
The FWHM of both distributions 
show that the energy deposition of minimum ionizing particles is measured with about $20-25~\%$ 
relative uncertainty. This is due to the low light yield of the current LSD module design.
Except for position 4, the values of the geometric mean respectively arithmetic mean at the maximum 
vary for the different beam positions by $10-15~\%$, which can be interpreted as a systematic uncertainty. 
These position-dependent differences are on average less pronounced for the geometric mean. 
One would expect equal geometric (arithmetic) mean values in particular 
for the positions 1, 2, 3, and 4, if there is no directional detection asymmetry of the two WOMs. 
While the geometric (arithmetic) mean values at the maximum agree within $10-15~\%$ for positions 
1, 2, and 3, they are significantly larger for position 4. This systematic difference may have been 
caused by liquid-scintillator leakage that occurred during the test beam measurements. This resulted 
in a small nitrogen-filled gap volume, leading to a liquid-scintillator surface in the upper part 
of the LSD module. This may result in total reflection of scintillation light close to the upper 
inner wall of the LSD (which is not possible for the other inner walls), thus affecting the light 
yield collection mainly for position 4.

In future measurements, all inner walls of the LSD will be covered by a reflective foil or coating to 
increase the light yield in general. Furthermore, a new LSD module will be produced from welded 
steel plates instead of glued ABS to avoid liquid scintillator leakage. An expansion vessel 
filled with liquid scintillator will be added that guarantees the LSD module to be always completely 
filled with liquid scintillator.\\

\begin{figure}
     \centering
     \subfigure[]{
          \label{fig:GeometricalMeanMuonsDifferentAngles}
          \includegraphics[scale=0.35]{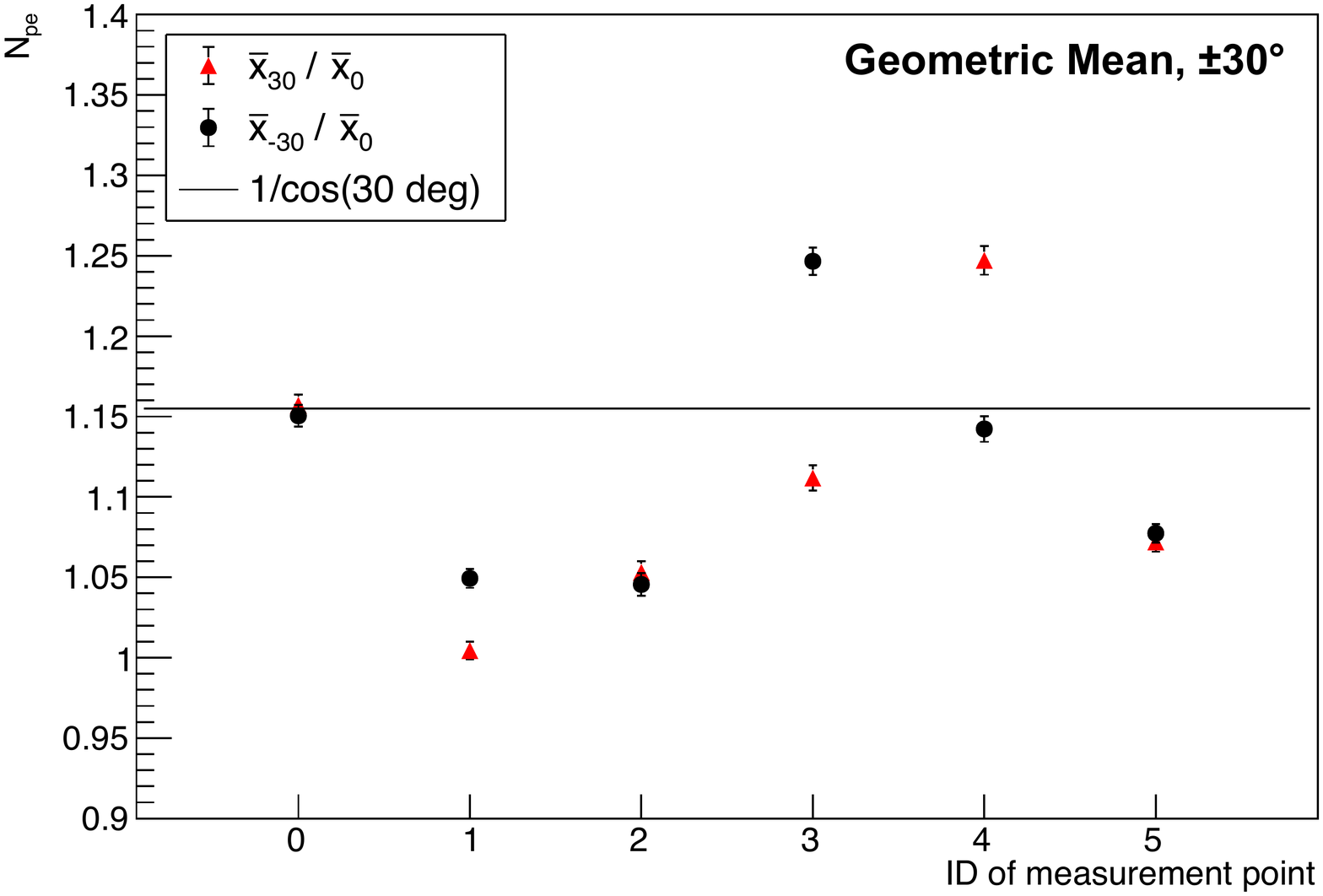}}
     \subfigure[]{
          \label{fig:MeanMuonsDifferentAngles}
          \includegraphics[scale=0.35]{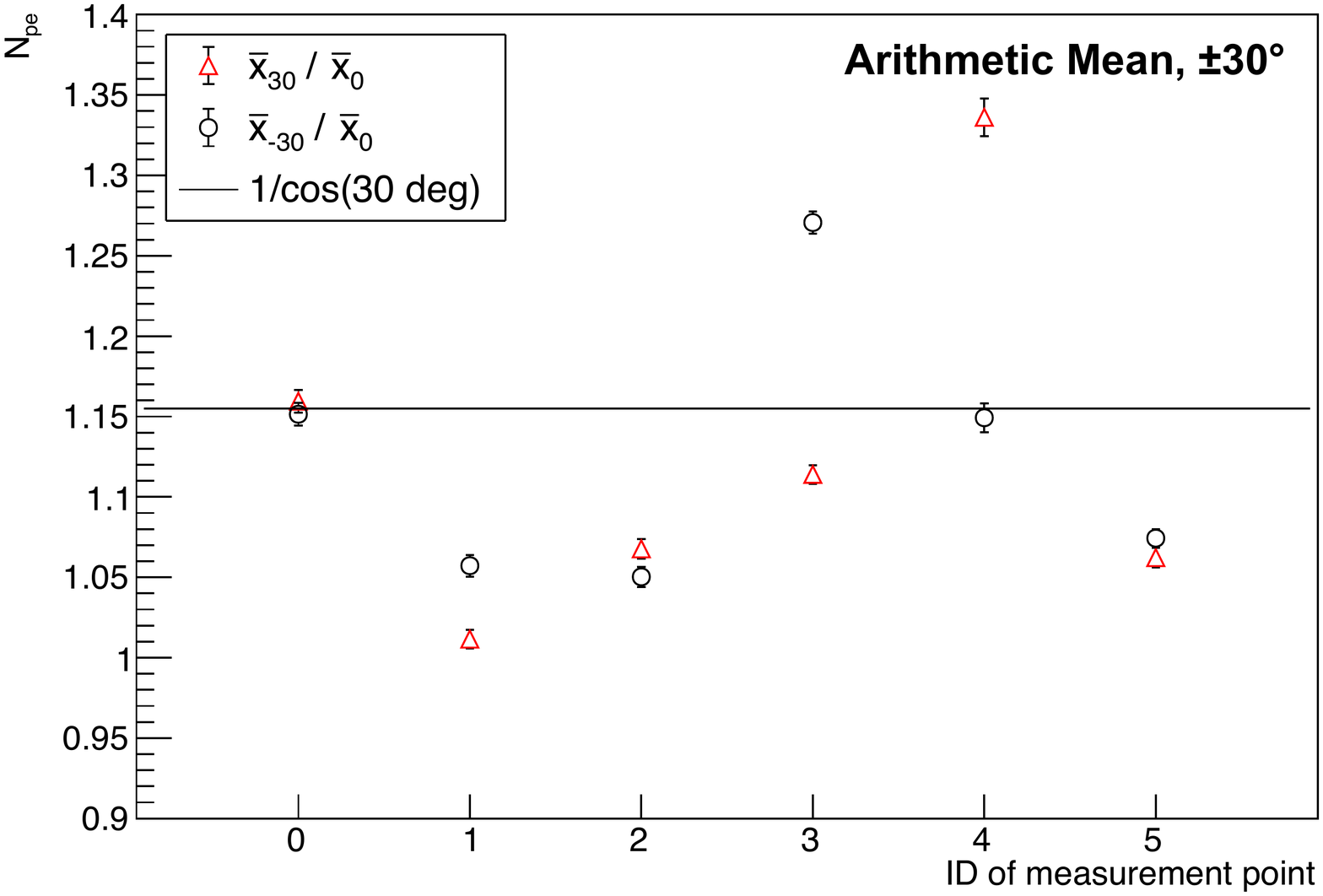}}
  \caption{The ratio of the observed photoelectron yield for incident angles of $+30^{\circ}$
           and $-30^{\circ}$ with respect to $0^{\circ}$ for  
           (a) geometric mean 
           and (b) arithmetic mean
           at different muon beam positions.}
  \label{fig:MeanGeometricalMeanMuonsDifferentAngles}
\end{figure}

For an incident angle different from $0^{\circ}$, one naively expects a $1/\cos{\theta_{inc}}$ higher 
energy deposition from a minimum ionizing particle, since the path length through the liquid 
scintillator increases by the same factor. 
Figs.~\ref{fig:GeometricalMeanMuonsDifferentAngles} and~\ref{fig:MeanMuonsDifferentAngles} show 
the ratio between the measurement at $\pm 30^{\circ}$ with respect to the measurement at $0^{\circ}$
for the geometric and arithemtic mean, respectively. 
On average, the results are in agreement with the expectation of $1/\cos{\theta_{inc}}=1.155$, 
however, significant deviations are observed. 
While for positions 0, 1, 2, and 5, the ratios for $+30^{\circ}$ respectively $-30^{\circ}$ are 
quite close to each other, they differ much more for positions 3 and 4, showing the opposite
behaviour.
This behaviour will be the subject of future more detailed studies and Monte Carlo simulations.
It might be e.g. partially caused by the WOM photo detection efficiency, which depends on 
the distance between the end of the WOM tube (where the photosensor is located) and the 
hit position of the primary scintillation photon.

\begin{figure}
  \begin{center}
             \includegraphics[scale=0.50]{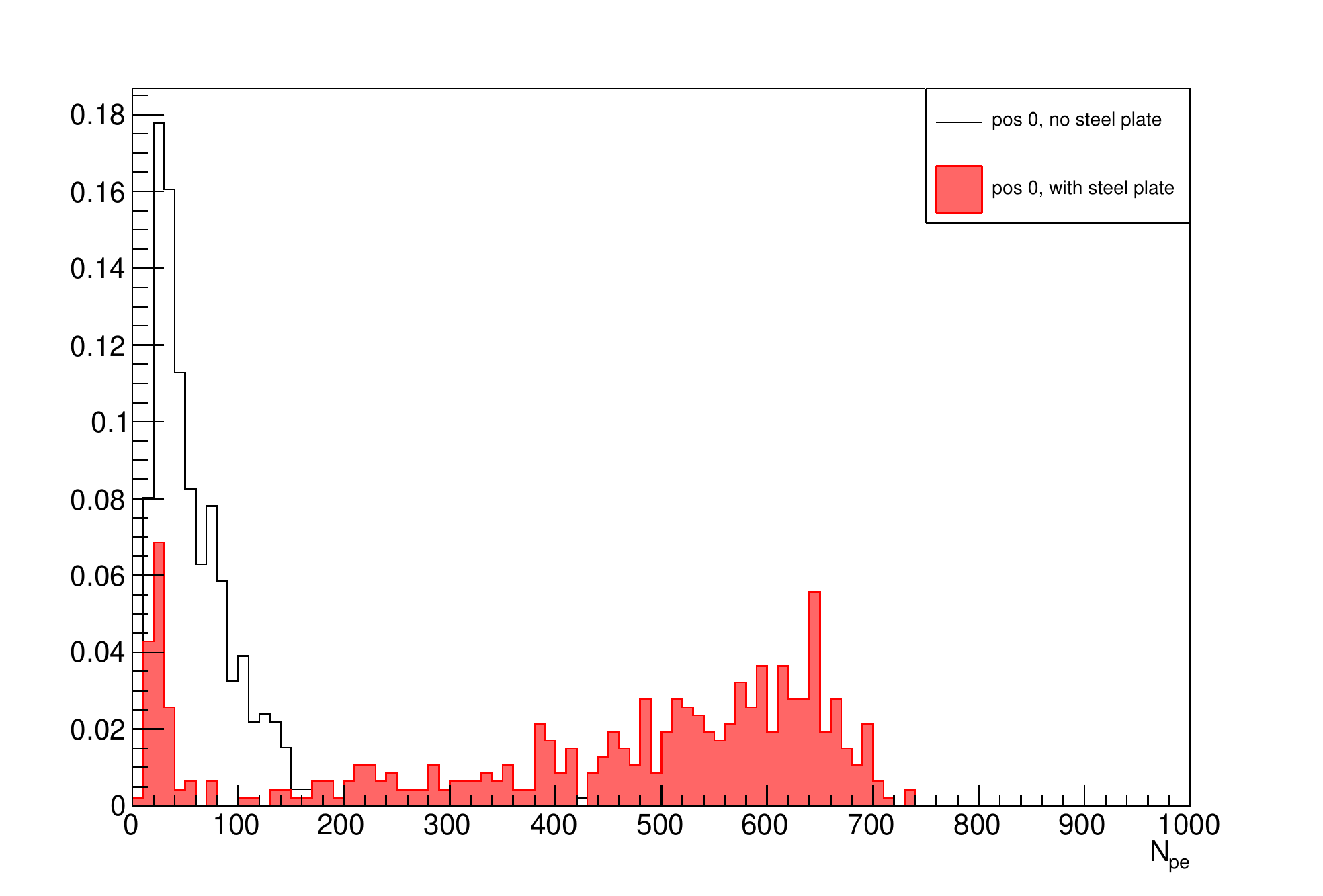}
  \end{center}
  \caption{$N_{\rm pe}$ distributions (normalized to the same integral)
           for the WOM with PMT2  
           for an electron beam at beam position 0 and an incident angle of 
           $0^{\circ}$ with and without a steel plate in front of the LSD module.}
  \label{fig:PMT_electrons}
\end{figure}

The specific response to electrons was studied with a steel plate placed in front of the LSD module, 
since the LSD can serve as a preshower detector to distinguish between electrons/photons and other 
particles. 
The radiation length of the steel plate was about $2.1 \cdot X_{0}$, with $X_{0}=1.76~{\rm cm}$ 
being the radiation length of stainless steel~\cite{PDG}. For a perpendicularly crossing particle, 
the liquid scintillator adds a a thickness of $O(0.6 \cdot X_{0})$.
This will produce an electromagnetic preshower with an expected energy deposition of $O(4~\%)$ 
of the incident electron energy~\cite{EMShower}, corresponding to about 1~GeV for the 20~GeV 
electrons used in the test-beam measurement. From the Bethe-Bloch formula, a minimum ionizing 
particle is expected to lose about 45~MeV energy on average when trasversing the liquid scintillator. 
For 20~GeV electrons, a factor of 20 more light yield is expected.
Fig.~\ref{fig:PMT_electrons} shows the $N_{\rm pe}^{PMT2}$ distributions for 20~GeV electrons 
hitting the LSD module at beam position 0 with an incident angle of $0^{\circ}$, with and without 
steel plate. The electron beam contamination from hadrons and muons is expected 
to be $O(20~\%)$. This is reflected by the observed peak at $N_{\rm pe} \approx 25$, corresponding to 
the expected 45~MeV of average energy deposition from minimum ionizing particles. The position of 
this peak is consistent with the results from a muon beam, see Fig.~\ref{fig:PE_Distribution_PMT}. 
A second broad peak $N_{\rm pe} \approx 600$ for PMT2 with the steel plate in front of the LSD module
is observed. Assumining linearity in the detector response, this is in agreement with the estimated 
energy deposition from electromagnetic preshowering induced by 20~GeV electrons in the steel plate. 
The fraction of events under the first peak from minimum ionizing particles accounts for about 
$O(20~\%)$ of the events, being consistent with the $O(80~\%)$ purity of the electron beam. 
The effect from the preshowering is heavily reduced when no steel plate is present. We conclude 
that the LSD can be used as a preshower detector for particle identifcation in connection with a 
sufficiently thick metal layer in front of the detector.

\section{Conclusion}
A proof-of-concept building large-volume liquid-scintillator detectors was conducted,
encompassing: wavelength-shifting optical modules for the collection of scintillation photons 
in the UV range, conversion of the UV photons into photons of longer wavelength in the WOM,
their detection by a vacuum photomultiplier tube or a SiPM array mounted on a light guide 
on one side of the WOM.

With a liquid-scintillator detector module of $50 \times 50 \times 30~{\rm cm}^3$ volume
and one WOM read out, photo detection efficiencies of at least $99.7~\%$ for a threshold 
of five photoelectrons are achieved. Using two WOMs, placed on opposite sides of the detector, 
results in a time resolution of about 1~ns. For the size of the detector module and obtained
average light yield, the effect on the time resolution from different positions of the 
traversing particle is found to be subdominant. 
The detector light yield for the combined signal of two WOMs on opposite sides of the 
detector is homogenous within $O(15\%)$ for particles perpendicularly traversing the detector. 
A layer of stainless steel in front of the detector demonstrates its capability to measure 
the preshowers from electrons providing much higher light yield than for muons and hadrons.\\
 
The light yield, photo detection efficiency, time resolution and possibly also the homogenity 
of the detector can still be improved in the future in several ways: increase of the PPO 
concentration in the liquid scintillator, purification of the LAB to obtain larger absorption 
lengths, coverage of all inner walls with material of good reflection properties for UV light, 
complete filling of the LSD (avoiding a liquid-scintillator surface leading to total reflection 
only at this surface), choosing PMMA material with higher transmission in the emission range of the PPO. 
Furthermore, the light guide and its coupling to the WOM tube can be improved, the coverage 
of the light guide exit by the photosensor can be optimized. As an alternative, SiPMs can be 
directly placed on the WOM tube without the need for any light guide. In this context, one
might profit from SiPMs of higher efficiency, such as the new Hamamatsu S14520 series.

\section*{ACKNOWLEDGEMENTS}
We acknowledge the support of I. Korol and P. Venkova by the DFG. The Geneva group acknowledges support 
from the Swiss National Science Foundation (grant PP00P2\_150583). We thank: Dustin Hebeker, Timo Karg, 
and Marek Kowalski (DESY) for the scientific discussions on using WOMs and providing the dip-coater for 
our WOM production; Peter Peiffer (Johannes Gutenberg-Universit\"at Mainz) for measuring the WOM 
photo detection efficiency with and without light guide; Steffen Hackbarth (Humboldt-Universit\"at zu Berlin) 
for measuring the transmission of the liquid scintillator in the UV range; Yannick Favre (University of Geneva) 
for design and assembling of SiPM boards; Vadim Denysenko and Oleg Bezshyyko
(Taras Shevchenko National University of Kyiv) for GEANT4 simulations of the WOM tube and light guide; 
Sergio G\'{o}mez and David Gasc\'{o}n (University of Barcelona) for support on the usage of the MUSIC board; 
Dominique Breton and Jihane Maalmi (LAL Orsay) for support on the usage of WaveCatcher. 
Special thanks go to Richard Jacobsson (CERN) and the SPS test-beam operators at CERN.

\end{document}